\documentclass[pdflatex,sn-mathphys-num]{sn-jnl}% Math and Physical Sciences Numbered Reference Style
\usepackage{float} 
\usepackage{graphicx}%
\usepackage{multirow}%
\usepackage{amsmath,amssymb,amsfonts}%
\usepackage{amsthm}%
\usepackage{mathrsfs}%
\usepackage[title]{appendix}%
\usepackage{xcolor}%
\usepackage{textcomp}%
\usepackage{manyfoot}%
\usepackage{booktabs}%
\usepackage{algorithm}%
\usepackage{algorithmicx}%
\usepackage{algpseudocode}%
\usepackage{listings}%

\theoremstyle{thmstyleone}%
\theoremstyle{thmstyletwo}%

\theoremstyle{thmstylethree}%

\begin{document}

\title[Article Title]{Random Exploration is Random}

%%=============================================================%%
%% GivenName	-> \fnm{Joergen W.}
%% Particle	-> \spfx{van der} -> surname prefix
%% FamilyName	-> \sur{Ploeg}
%% Suffix	-> \sfx{IV}
%% \author*[1,2]{\fnm{Joergen W.} \spfx{van der} \sur{Ploeg} 
%%  \sfx{IV}}\email{iauthor@gmail.com}
%%=============================================================%%

\author*[1, 2]{\fnm{Yingkang} \sur{Luo}}
\email{yluo344@gatech.edu}

\author[2]{\fnm{Hua-Dong} \sur{Xiong}}
\email{hdx@gatech.edu}

\author[2]{\fnm{Hanbo} \sur{Xie}}
\email{hanboxie1997@gatech.edu}

\author[2]{\fnm{Caroline} \sur{Phelps}}
\email{cphelps35@gatech.edu}

\author[3]{\fnm{Madeline} \sur{Harms}}
\email{madeline.harms@gmail.com}

\author[4]{\fnm{Ryan} \sur{Smith}}
\email{rsmith@laureateinstitute.org}

\author[5]{\fnm{James} \sur{Waltz}}
\email{JWaltz@som.umaryland.edu}

\author*[2]{\fnm{Robert} \sur{Wilson}}
\email{rwilson337@gatech.edu}

% reach out to authors of papers with dataset 

\affil[1]{\orgdiv{College of Computing}, \orgname{Georgia Institute of Technology}
% \orgaddress{\street{225 North Ave}, \city{Atlanta}, \postcode{30332-0530}, \state{Georgia}, \country{U.S.A.}}
}

\affil[2]{\orgdiv{School of Psychology}, \orgname{Georgia Institute of Technology}
% \orgaddress{\street{225 North Ave}, \city{Atlanta}, \postcode{30332-0530}, \state{Georgia}, \country{U.S.A.}}
}

\affil[3]{\orgdiv{Department of Psychology}, \orgname{University of Minnesota Duluth}
% \orgaddress{\street{225 North Ave}, \city{Atlanta}, \postcode{30332-0530}, \state{Georgia}, \country{U.S.A.}}
}
\affil[4]{\orgname{Laureate Institute for Brain Research}
% \affil[4]{\orgdiv{Community Medicine}, \orgname{University of Tulsa}
% \orgaddress{\street{225 North Ave}, \city{Atlanta}, \postcode{30332-0530}, \state{Georgia}, \country{U.S.A.}}
}

\affil[5]{\orgdiv{School of Medicine}, \orgname{University of Maryland}
% \orgaddress{\street{225 North Ave}, \city{Atlanta}, \postcode{30332-0530}, \state{Georgia}, \country{U.S.A.}}
}

%%==================================%%
%% Sample for unstructured abstract %%
%%==================================%%

\abstract{Decision making often requires balancing exploitation of familiar options with exploration of uncertain alternatives. One strategy for exploration is random exploration, in which choices become more variable and less tied to the option that currently appears best. However, it remains unclear whether this variability reflects genuine stochasticity or instead a deterministic structure that traditional cognitive models fail to capture. We examined this question using the Horizon Task in an aggregated dataset of 1,392 participants completing over 1,270,695 trials. In this task, exploration becomes more valuable when participants have more future choices, and choice variability increases under longer horizons. We tested whether more expressive neural network models could recover deterministic structure from this additional variability. Although neural networks provided better overall fits to behavior than the traditional model, long-horizon choices remained consistently less predictable than short-horizon choices. These findings show that limited model expressiveness cannot fully account for the increased variability associated with random exploration, suggesting random exploration truly is substantially random.}

\keywords{Random Exploration, Neural Networks, Explore-exploit, Decision-making}

%%\pacs[JEL Classification]{D8, H51}

%%\pacs[MSC Classification]{35A01, 65L10, 65L12, 65L20, 65L70}

\maketitle
\section{Introduction}\label{sec1}
When dining out, do you order a familiar favorite or try something different? 
You can't go wrong with a steak, but the sea urchin stir-fry might taste even better. There are also spontaneous moments when we act without quite knowing what is driving us: the server arrives, and suddenly we hear ourselves ordering the fish-head soup! Every time we make a decision, we are facing the explore-exploit dilemma: choosing between exploiting what we already know and exploring uncertain alternatives that may improve future decisions \cite{Wilson2014, Feng2021, Sutton1998, kaelbling1996reinforcementlearningsurvey}. It can be as simple as deciding what to eat or as complex as choosing a profession. Exploiting a familiar option is safe and predictable, but it may mean missing out on potentially better alternatives. Trying something new may reveal a better option, but it also comes with the risk of disappointment. 

Additionally, the value of exploration often depends on how many future opportunities remain. Imagine that you are staying in a new city for several months. Trying an unfamiliar dish may help you discover a favorite that you can return to many times. If you are leaving tomorrow, however, there may be little opportunity to benefit from what you learn. This number of remaining opportunities is known as the horizon. When horizon is long, new information becomes more valuable, encouraging exploration. Conversely, when future opportunities are limited, exploitation becomes more favorable \cite{bellman1956problem, gittins1979bandit, zhang2013forgetful, averbeck2015theory}.

% \subsection{Directed and Random Exploration}
Previous work has identified two strategies for navigating the explore-exploit dilemma: directed and random exploration. Directed exploration involves deliberately favoring more informative options. Random exploration involves noisier choices that are less tied to expected value or the degree of uncertainty \cite{bellman1956problem, gittins1979bandit, Wilson2021, krebs1978test, watkins1989learning, gershman2018dopaminergic, schulz2019algorithmic, gershman2018deconstructing, frank2009prefrontal, badre2012rostrolateral, cavanagh2012frontal, speekenbrink2015uncertainty, daw2006cortical}. 
This noise can draw an individual away from the option with the highest expected reward. Although introducing randomness, or stochasticity into decision-making might seem counterproductive, it provides an effective and computationally inexpensive way to explore. Stochastic methods are therefore widely used in machine-learning algorithms for search and optimization \cite{Kirkpatrick1983, Cerny1985, Geman1984, Metropolis1953, Hastings1970, Gelfand1990}. Consistently choosing the option with the highest expected value risks becoming trapped in a familiar but suboptimal pattern. Stochasticity disrupts this pattern and creates opportunities to discover better options. 

However, is behavioral variability commonly attributed to random noise truly random, or does it reflect structured processes that commonly used models fail to identify?
Evidence from neuroscience suggests that genuine neural noise may contribute to random exploration. Neural activity is inherently variable, and this variability can produce stochastic behavior \cite{faisal2008noise}. During random exploration, monkeys show increased neural variability and weaker choice-related signals in prefrontal decision-making circuits \cite{Ebitz2018}. At the same time, random exploration appears to be strategically regulated. Like directed exploration, it increases when exploration is more valuable, and studies in rodents and songbirds suggest that this may involve increased neural variability \cite{Tervo2014,Budzillo2017}.
Together, these findings suggest that apparently random behavior reflects a regulated process.

Furthermore, not all behavioral variability reflects irreducible randomness. Apparent randomness can also arise from noise in how task or action values are learned rather than from noise at the moment of choice \cite{Findling2019}. Sensory misestimation, imprecise learning, and simplified beliefs can produce variable behavior even when the individual is attempting to follow a value-maximizing strategy  \cite{osborne2005sensory,Beck2012}. Wang and colleagues examined this issue by modifying the Horizon Task to present participants with repeated identical games. The task compares choices under short (H1) and long (H6) horizons, where exploration becomes more valuable in H6. 
Some choice variability was consistent across repetitions, while a substantial portion remained unexplained, supporting the possibility of a genuinely stochastic component \cite{Wang2026}. However, repeating the same game only controls the choices and rewards within that game. It does not account for influences from earlier games, gradual changes in strategy, or other unobserved factors \cite{xiong2025strategy}. Thus, variability that appears random may still reflect deterministic processes that the model has not captured. 

In practice, randomness is identified as variance in behavior that remains unexplained after a model has accounted for the variables of interest. Traditional cognitive models often represent the unexplained choice variability through a decision-noise parameter \cite{Wilson2014, gershman2019uncertainty}. However, these simplified assumptions restrict the behavioral relationships they can capture. Some apparently random variability may therefore be systematic but unmodeled, making the model's expressive power a key limitation.

The current study addresses this problem using deep neural networks in previously collected Horizon Task data with over one million trials \cite{Harms2024, Phelps2025, Mizell2024, Sadeghiyeh2020, Smith2022, Somerville2017, Waltz2020, Wilson2014}. Neural networks are powerful function approximators with scalable numbers of trainable parameters and flexible architectures \cite{lecun2015deep, hornik1991approximation, Cybenko1989}. This allows them to capture complex behavioral patterns that traditional cognitive models may have misclassified as randomness. In the Horizon Task, we observed greater behavioral variability at longer horizons, which was attributed as decision noise by the traditional model. If this variability contains additional predictable structure, neural networks should improve prediction more at long- than short-horizon games, reducing the predictability gap between conditions. If long-horizon choices remain less predictable, limited model expressiveness cannot fully explain the additional variability, supporting a genuinely stochastic component of random exploration.

\section{Results}\label{sec2}
Before presenting the results, we briefly describe the Horizon Task and provide an overview of our modeling approach.

\subsection{The Horizon Task}

In the Horizon Task \cite{Wilson2014}, participants play a series of games in which they choose between two slot machines to maximize their total reward. Each machine produces rewards from a Gaussian distribution with a different mean unknown to the participants. They must therefore balance exploiting the option with the highest expected reward against exploring to gain additional information.

\begin{figure*}[p]
\centering
  \includegraphics[width=1\columnwidth]{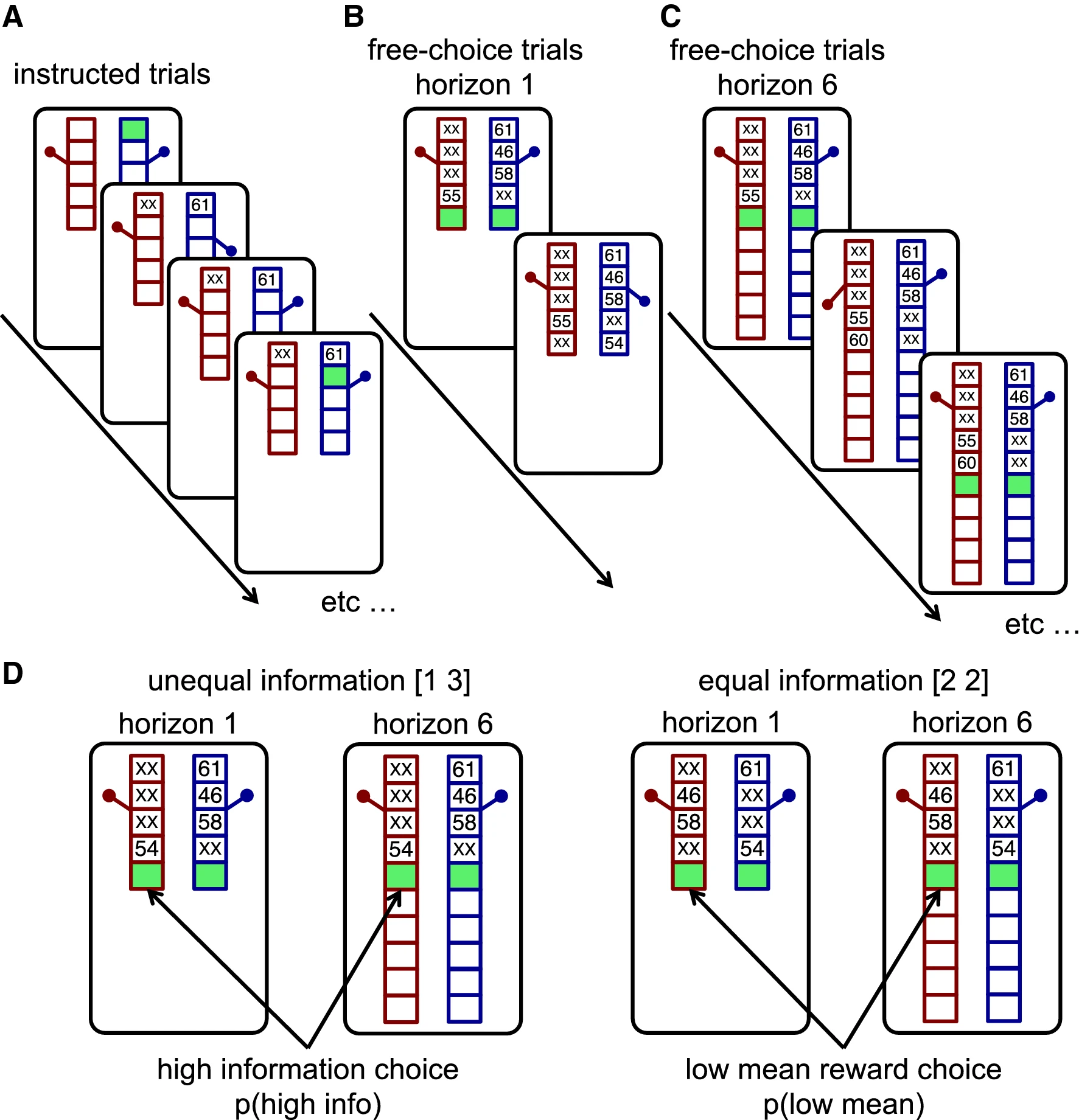}
  \caption{The Horizon Task. (A) The first four instructed trials, during which participants are required to select the highlighted green option between the two slot machines. The structure of the forced trials is identical across horizon conditions, differing only in the total number of trials. (B, C) Following the instructed trials, participants complete free-choice trials until the end of the game. During this phase, they choose freely between the two slot machines. In the short-horizon condition (B), participants make one additional free choice, whereas in the long-horizon condition (C), they make six additional free choices. (D) Illustration of the uncertainty conditions. On the left, the instructed trials reveal an unequal amount of information per machine [1 3]. On the right, the instructed trials reveal an equal amount of information per machine [2 2]. Analyses focus on the first free-choice trial, highlighted in green.}~\label{fig:horizon}
\end{figure*}
\clearpage

Each game begins with four instructed trials, during which participants are required to choose a specified machine (Fig. \ref{fig:horizon}). These trials control how much information is available about each option before participants make a free choice. In the unequal uncertainty condition, one machine is played once and the other three times ([1 3]). In the equal uncertainty condition, both machines are played twice ([2 2]). After the instructed trials, participants make either one free choice in Horizon 1 (H1) or six free choices in Horizon 6 (H6). In H1, information gained through exploration cannot be used in a later choice. In H6, that information may improve future decisions, making exploration more valuable. We focus primarily on the first free-choice trial, where the available choice and reward information is controlled in the same way across horizon conditions. Comparing behavior between H1 and H6 on this trial allows us to measure horizon-related changes in behavioral variability.

\subsection{Modeling Overview}
We analyzed an aggregated Horizon Task dataset of 1,392 participants. We used the original Horizon Task model as a baseline \cite{Wilson2014} and confirmed that the dataset reproduced the established effects of horizon on directed and random exploration. We then used neural networks (NNs) to capture additional predictable structure available in the task. To reduce model-specific limitations, we evaluated several architectures with different representational capacities. The multilayer perceptron (MLP) received all trial information simultaneously, whereas recurrent neural networks (RNNs) processed choices and rewards sequentially, similar to how information was presented to participants. We also evaluated Transformer-based models to capture flexible dependencies across trials.

We first tested whether the neural networks captured more predictable behavioral structure than the traditional cognitive model. Then, we examined whether this additional predictive capacity reduced the H1–H6 predictability gap across several complementary analyses. In the simplest case, we modeled behavior at the pooled-population level using only the instructed-trial choice–reward sequence, horizon condition, and uncertainty condition. We then incorporated subject identity, allowing the models to learn individual behavioral patterns. To rule out learning effects from prior games, we repeated the analysis using only the first game completed by each participant. We next tested for cross-game history effects using RNNs with a learned persistence parameter that carried information forward from previous games within each participant. Finally, we tracked predictive performance across all six free-choice trials in H6 to examine how predictability changed as the remaining horizon shortened.

\subsection{Unexplained variance of the traditional cognitive model increases with horizon}

\begin{figure*}[p]
\centering
    \includegraphics[
        width=\textwidth,
        height=0.9\textheight,
        keepaspectratio
    ]{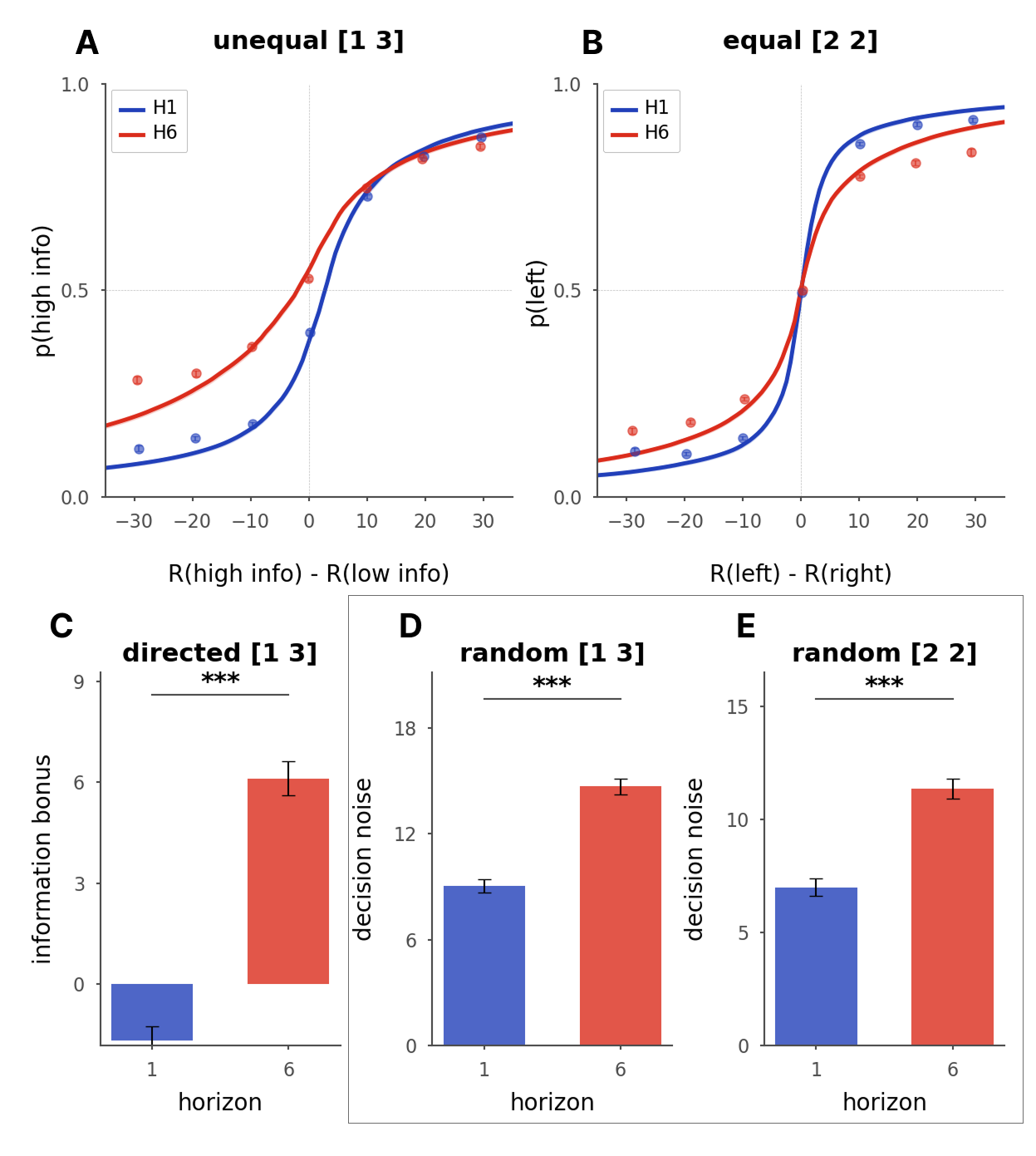}
    \caption{Choice curves of the Horizon Task. (A, B) Choice curves showing the probability of selecting an option as a function of the difference in observed mean rewards. (A) Probability of choosing the more informative option in the unequal uncertainty condition (i.e., the machine played once during the forced trials). The observed mean reward difference is the reward of the more informative option minus that of the less informative option. (B) Probability of choosing the left option in equal condition as a function of the observed mean reward difference between the left and right options. (C, D, E) Parameter estimates from logistic regression fits across horizon conditions. (C) Information bonus in the unequal uncertainty condition. (D, E) Decision noise parameters in the unequal and equal uncertainty conditions, respectively. *** denotes a significant difference between horizon conditions at $p < 0.001$.}~\label{fig:original}
\end{figure*}

We first fitted the standard Horizon Task logistic model to establish a baseline for behavioral variability explained by the model. Consistent with previous findings, participants showed greater information seeking in H6 than in H1 (Fig.~\ref{fig:original}A). This was reflected in a larger information bonus in the long-horizon condition ($t(1389)=-15.74$, $p<.001$) (Fig.~\ref{fig:original}C).

\clearpage

Additionally, the model attributed more behavioral variability to decision noise as horizon increased (Fig.~\ref{fig:original}B). Choices were more variable in H6 than in H1, and the fitted decision-noise parameter was higher in H6 in both uncertainty conditions (Fig.~\ref{fig:original}D,E). Decision noise increased significantly in the equal uncertainty condition ([2 2]: $t(1389)=-14.97$, $p<.001$) and in the unequal uncertainty condition ([1 3]: $t(1267)=-12.25$, $p<.001$).

To mirror the neural network analyses, we also evaluated the logistic model as an out-of-sample predictive baseline under different fitting procedures (Table~\ref{tab:baseline_accuracy}). The global fit estimated a single set of parameters across the pooled population, similar to the population-level fitting commonly used in neural network models \cite{Peterson2021}. The individual fit instead estimated a separate set of logistic-model parameters for each participant, which were then averaged for group-level summaries. Under both procedures, H6 choices were harder to predict than H1 choices. The H1--H6 accuracy difference was 7.5 percentage points for the global fit and 6.1 percentage points for the individual fits. We also evaluated the logistic model using only the first game completed by each participant. The predictability gap persisted, with an accuracy of 0.881 in H1 compared with 0.767 in H6. Thus, the lower predictability of H6 behavior was present before participants accumulated experience across repeated games.
Overall, the aggregated dataset reproduced the established Horizon Task effects and showed that long-horizon behavior was less predictable across fitting procedures and from the first game.

\begin{table}[h!]
\caption{Horizon-specific test accuracy of the logistic regression baselines.}
\label{tab:baseline_accuracy}
\begin{tabular*}{\textwidth}{@{\extracolsep\fill}lccc}
\toprule
Baseline fitting procedure & H1 accuracy & H6 accuracy & H1--H6 difference \\
\midrule
Global fit      & 0.807 & 0.732 & 0.075 \\
Individual fits & 0.799 & 0.737 & 0.061 \\
First-game fit  & 0.881 & 0.767 & 0.113 \\
\botrule
\end{tabular*}
\end{table}

\subsection{Neural network models fit better than the baseline cognitive model}

The traditional Horizon Task model has only a small number of free parameters, which limits the range of behavioral structure it can capture. As a result, deterministic patterns that fall outside the model may be absorbed into the decision-noise term and appear stochastic. We therefore tested whether more expressive neural network models provided a better fit to behavior. The neural network models captured more behavioral structure than the baseline logistic model. We compared model fit using McFadden's pseudo-$R^2$ relative to the null model (Table~\ref{tab:nll}). The logistic regression baseline achieved a pseudo-$R^2$ of $0.217$. In contrast, all neural network models from the subject-level analysis achieved higher values, ranging from $0.312$ to $0.385$. Overall, the neural network models increased McFadden's pseudo-$R^2$ by 0.095--0.168 relative to the logistic regression baseline.

\begin{table}[h!]
\caption{Model comparison using NLL and McFadden's pseudo-$R^2$.}
\label{tab:nll}
\begin{tabular*}{\textwidth}{@{\extracolsep\fill}lccc}
\toprule
Model & NLL & Pseudo-$R^2$ & $\Delta R^2$ vs.\ Baseline \\
\midrule
MLP             & 0.4440 & 0.3593 & 0.1426 \\
RNN             & 0.4604 & 0.3356 & 0.1189 \\
LSTM            & 0.4282 & 0.3821 & 0.1655 \\
GRU             & 0.4350 & 0.3722 & 0.1556 \\
TinyGRU         & 0.4769 & 0.3118 & 0.0952 \\
Transformer     & 0.4323 & 0.3761 & 0.1595 \\
Self-Attention  & \textbf{0.4262} & \textbf{0.3850} & \textbf{0.1684} \\
Baseline        & 0.5428 & 0.2167 & 0.0000 \\
\botrule
\end{tabular*}
\end{table}

\subsection{Unexplained variance of all neural network models increases with horizon}
\begin{figure*}[p]
\centering
    \includegraphics[
        width=\textwidth,
        height=0.82\textheight,
        keepaspectratio
    ]{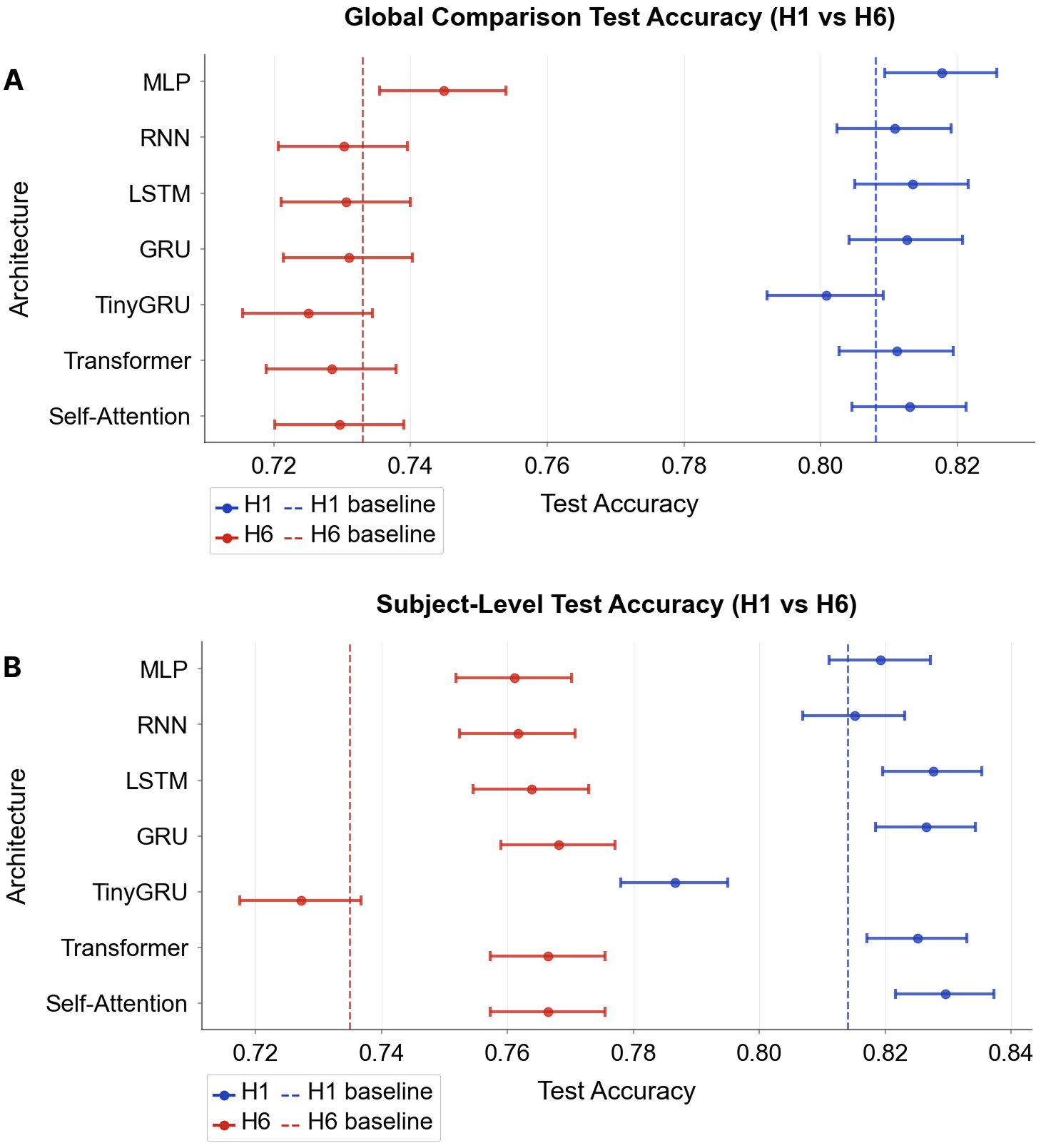}
    \caption{Test accuracy by horizon across neural network architectures. (A) Global analysis, where models predicted choices without participant identity. (B) Subject-level analysis, where participant identity was included as an additional input. Each point shows the test accuracy of one architecture, computed separately for H1 and H6 after restoring the checkpoint with the lowest validation loss. Horizontal error bars indicate 95\% Wilson confidence intervals. H1 accuracy is shown in blue and H6 accuracy in red. Dashed vertical lines indicate the corresponding horizon-specific test accuracy of the logistic regression baseline. Abbreviations: MLP, multilayer perceptron; RNN, recurrent neural network; LSTM, long short-term memory; GRU, gated recurrent unit; H1, Horizon 1; H6, Horizon 6.}
    
    \label{fig:all}
\end{figure*}

We first tested the simplest case by training neural network models on the pooled population without subject identity. These models predicted how a typical participant in the pooled population would choose. Across all architectures, test accuracy was higher in H1 than in H6 (Fig.~\ref{fig:all} A). Mean H1 accuracy was $0.811$ (95\% CI: $[0.808, 0.815]$), compared with $0.731$ for H6 (95\% CI: $[0.727, 0.736]$). H1 accuracy was therefore $8.0$ percentage points higher than H6 accuracy. Thus, the global neural network models did not close the predictability gap between horizon conditions. H6 choices remained less predictable than H1 choices, consistent with greater exploratory variability in the long-horizon condition.

% \begin{figure}[H]
% \centering
%   \includegraphics[width=1\columnwidth]{figures/all analyses.png}
%   \caption{Predictability difference overview between H1 and H6 across neural network analyses. Test accuracy was compared between H1 and H6 across the global, subject-level, cross-game, and first-game analyses. Each point shows the mean test accuracy across all neural network architectures within a given analysis, computed separately for the H1 and H6 test subsets. Horizontal error bars indicate the 95\% confidence interval around the mean across architectures. H1 accuracy is shown in blue and H6 accuracy is shown in red. Dashed vertical lines indicate the horizon-specific accuracy of the logistic regression baseline matched to each analysis. Each baseline was fitted using the same data scope and level of analysis as the corresponding neural network models to ensure a fair comparison.}~\label{fig:all_analyses}
% \end{figure}

Adding subject identity allowed us to examine whether individual behavioral patterns accounted for the H1--H6 predictability gap. Test accuracy remained higher in H1 than in H6 across all neural network models with encoded subject identity information (Fig.~\ref{fig:all} B). Across models, mean H1 accuracy was $0.819$ (95\% CI: $[0.807, 0.830]$), compared with $0.759$ for H6 (95\% CI: $[0.749, 0.770]$). H1 accuracy was therefore $6.0$ percentage points higher than H6 accuracy. Thus, encoding subject identity improved mean predictive accuracy relative to the logistic baseline but did not eliminate the H1--H6 predictability gap.

\clearpage

\subsection{Unexplained variance of the neural network models increases from the very first game}
\begin{figure}[H]
\centering
  \includegraphics[width=1\columnwidth]{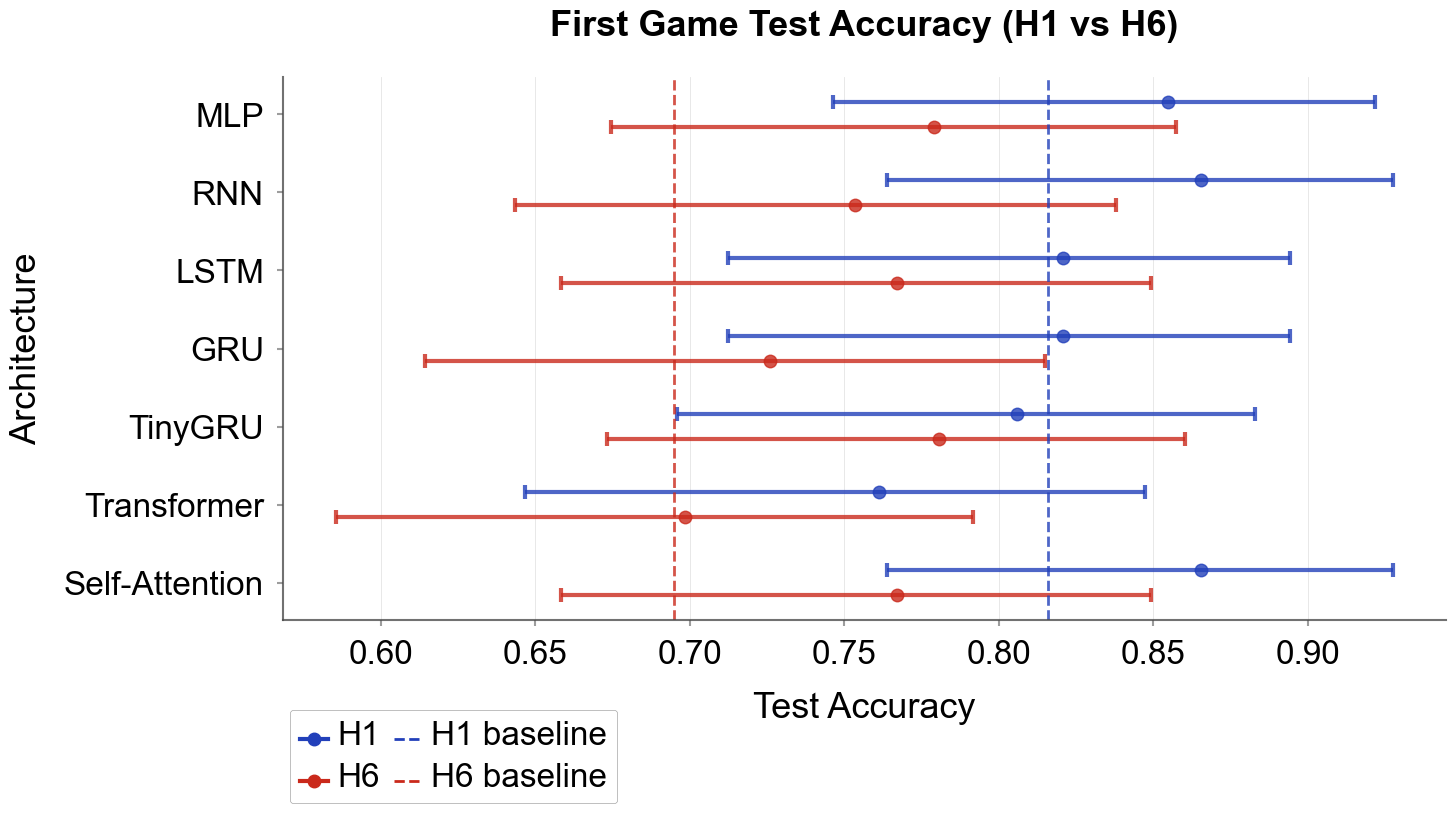}
\caption{Predictive performance in the first game. Test accuracy was compared between H1 and H6 using only the first game played by each participant. Each point shows the test accuracy of one neural network architecture after restoring the checkpoint with the lowest validation loss, computed separately for the H1 and H6 conditions. Horizontal error bars indicate 95\% Wilson confidence intervals for test accuracy. H1 accuracy is shown in blue and H6 accuracy is shown in red. Dashed vertical lines indicate the corresponding horizon-specific test accuracy of the first-game logistic regression baseline. Abbreviations: MLP, multilayer perceptron; RNN, recurrent neural network; LSTM, long short-term memory; GRU, gated recurrent unit; H1, Horizon 1; H6, Horizon 6.}

\label{fig:first}

\end{figure}

Behavior may change as participants gain experience with the task, raising the possibility that the H1--H6 predictability gap develops through learning or strategy adaptation across repeated games \cite{xiong2025strategy}. To test this possibility, we repeated the analysis using only the first game played by each participant. Test accuracy was higher in H1 than in H6 across all neural network architectures (Fig.~\ref{fig:first}). Mean H1 accuracy was $0.828$ (95\% CI: $[0.800, 0.856]$), compared with $0.753$ for H6 (95\% CI: $[0.731, 0.776]$), corresponding to an H1--H6 difference of $7.5$ percentage points. The predictability gap was therefore already present during participants' first exposure to the task, suggesting that it did not arise solely from learning or adaptation across repeated games.

\subsection{Unexplained variance is not accounted for by history effects from game to game}
Previous analyses captured dependencies between trials within the current game, but behavior may also depend on experiences from earlier games. These cross-game history effects provide another possible source of deterministic structure. Repeating the same game controls the choices and rewards within that game, but does not control for influences carried over from previous games \cite{Wang2026}. To test for these effects, we used RNNs that propagated their hidden state from one game to the next within each participant. A learned parameter $\beta$ controlled how strongly information from the previous game influenced the current game.

\begin{figure}[H]
\centering
  \includegraphics[width=1\columnwidth]{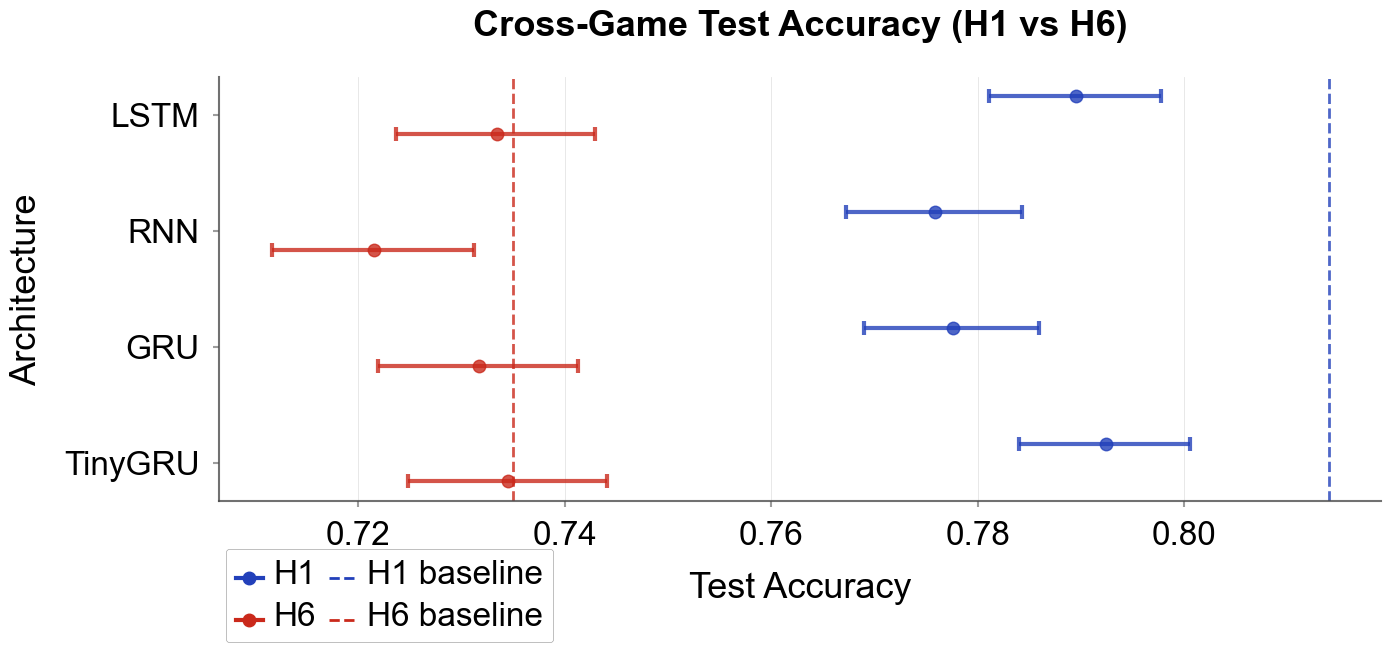}
\caption{Cross-game history analysis. Predictive accuracy was compared between H1 and H6 for recurrent models that incorporated cross-game history through the persistence parameter $\beta$. Each point shows the test accuracy of one architecture after restoring the checkpoint with the lowest validation loss, computed separately for the H1 and H6 conditions. Horizontal error bars indicate 95\% Wilson confidence intervals for test accuracy. H1 accuracy is shown in blue and H6 accuracy is shown in red. Dashed vertical lines indicate the corresponding horizon-specific test accuracy of the individual-fit logistic regression baseline. Abbreviations: RNN, recurrent neural network; LSTM, long short-term memory; GRU, gated recurrent unit; H1, Horizon 1; H6, Horizon 6.}

\label{fig:crossgame}

\end{figure}

Including cross-game history did not eliminate the predictability gap between H1 and H6 (Fig.~\ref{fig:crossgame}). Across models, mean H1 accuracy was $0.784$ (95\% CI: $[0.776, 0.792]$), compared with $0.730$ for H6 (95\% CI: $[0.724, 0.736]$), corresponding to an H1--H6 difference of $5.4$ percentage points. The models also learned relatively low $\beta$ values (LSTM $= 0.139$, RNN $= 0.183$, GRU $= 0.245$, TinyGRU $= 0.019$), indicating that prediction relied more heavily on information from the current game than on information carried over from previous games. Thus, history from prior games’ instructed phases provided little additional explanation for the lower predictability of H6 choices.

\subsection{Unexplained variance decreases over the course of horizon 6 games}

\begin{figure}[H]
\centering
  \includegraphics[width=1\columnwidth]{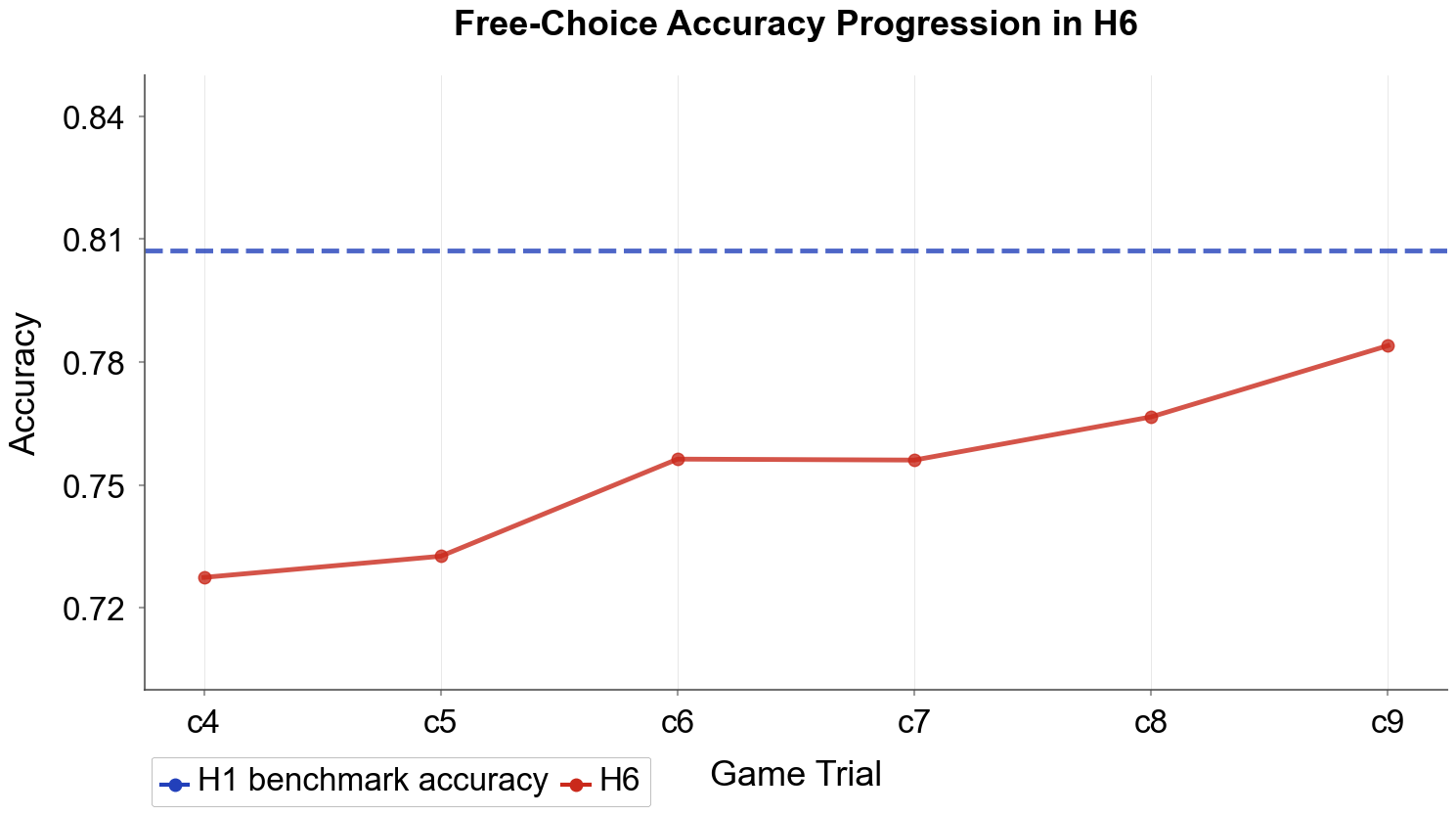}
  \caption{Predictive accuracy across H6 free-choice trials. The red line shows Seq2Seq model accuracy across the six free-choice trials in the H6 condition, from $C_4$ to $C_9$. The dashed blue line indicates the H1 benchmark accuracy. Predictive accuracy increased across later H6 trials as the remaining horizon shortened, but final-trial H6 performance remained below the H1 benchmark.}~\label{fig:later}
\end{figure}

If the increased unpredictability in H6 reflects the value of future exploration, it should decrease as the remaining horizon shortens within a game. We therefore examined how predictive accuracy changed across the six free-choice trials in H6. As the game progressed, choices became increasingly predictable (Fig.~\ref{fig:later}). Accuracy was lowest in the early H6 trials and generally increased across later trials. However, even by the final free-choice trial, H6 accuracy remained below the H1 benchmark. Overall, predictability increased as the remaining horizon shortened, consistent with greater behavioral variability when future exploration was more valuable.

\section{Discussion}\label{sec12}
We investigated whether behavior previously described as random exploration could instead be explained by a deterministic structure that simpler cognitive models fail to capture. Neural network models captured more behavioral structure than the traditional logistic model, suggesting that some predictable information was missed by the simpler model. However, this additional model capacity did not completely eliminate the lower predictability of H6 choices. The reported H1–H6 accuracy gap narrowed in the subject-level and first-game comparisons, suggesting that the networks recovered some predictable structure in long-horizon choices. Nevertheless, H6 choices remained less predictable across the neural-network analyses, a residual difference that does not by itself establish irreducible stochasticity.

We tested several possible sources of deterministic structure. At first, the models used information from the current game, which allowed them to account for nonlinear relationships among the instructed choices, rewards, and task conditions, while recurrent and attention-based models could additionally capture sequential dependencies within the game. Nevertheless, H6 choices remained less predictable than H1 choices. We then included subject identity, which allowed the models to capture individual behavioral patterns and tested whether the H1–H6 predictability gap was an artifact of averaging across individuals. The predictive accuracy improved, but the H1--H6 gap remained. 

Moreover, we also tested whether the predictability difference depended on participants' experience with the task. The H1–H6 predictability gap was already present when the analysis was restricted to each participant's first game. We also tested whether experiences from previous games contributed to the gap by incorporating cross-game history into the models. However, this additional information still did not eliminate the predictability gap. Together, these findings suggest that the lower predictability of H6 choices did not arise solely from learning or strategy adaptation across repeated games. Within H6 games, choices became more predictable as the remaining horizon shortened, consistent with the increase in variability being linked to opportunities for future exploration. Together, these analyses show that the lower predictability of long-horizon choices persists after accounting for several sources of deterministic structure available in the Horizon Task.

A key limitation is that even highly expressive models can only capture deterministic structure that is present in the measured data. Our neural networks had access to choices, rewards, task conditions, subject identity, and, in some analyses, previous game history. However, other factors that may systematically influence choice were not measured. These could include where participants were looking, subtle movements or fidgeting, and internal states such as attention, arousal, or boredom. For example, spontaneous movements that are typically unrelated to the task can account for substantial trial-to-trial variation in neural activity \cite{Musall2019}. If similar unmeasured factors influence exploratory choices, their effects would appear as unexplained variability in our models even if they were deterministic.

Future work could address this limitation by measuring a richer set of behavioral and physiological signals alongside choice. Eye tracking could test whether gaze patterns change with horizon, while video or motion tracking could measure fidgeting and other spontaneous movements. Physiological measures, such as pupil dilation and heart rate, could similarly capture changes in internal state. Incorporating these features would provide a stronger test of how much of the remaining variability in long-horizon choices is recoverable.

A second limitation is that we prioritized predictive performance over cognitive constraints and interpretability. To obtain the best possible fit to behavior, we did not examine how the networks represented the behavioral structure they learned. Several approaches could make these models more interpretable. Machine-learning models can be constrained to produce explicit psychological theories \cite{Peterson2021}, while small recurrent networks can reveal compact latent strategies through analysis of their internal dynamics \cite{Ji-An2025}. Related work has also used structured cognitive models to separate interpretable components of complex decision behavior \cite{Wang2025FLAG}. Applying similar approaches here could reveal whether the neural networks recover known components of exploration, such as reward sensitivity, uncertainty, and information seeking, or identify additional behavioral states that are not represented in the traditional model.

Taken together, our findings suggest that the additional variability observed during long-horizon exploration cannot be explained by limited model expressiveness alone. More flexible models recovered substantially more predictable structure in behavior, yet long-horizon choices remained less predictable across the sources of information we examined. Within the behavioral measurements available in the Horizon Task, this residual variability is therefore consistent with a genuinely stochastic component of random exploration. Future work that incorporates richer behavioral and physiological measurements may help determine how much of this remaining variability can ultimately be explained.

\section{Methods}\label{sec11}

\subsection{The Horizon Task}
In the Horizon Task, participants choose between two slot machines, or one-armed bandits, with the goal of maximizing their total reward \cite{Wilson2014}. When selected, each bandit produces a reward sampled from a Gaussian distribution. Rewards are truncated to fall between 1 and 100 points and rounded to the nearest integer. The two bandits have different mean rewards, such that one option is better on average. The mean of one bandit is set to either 40 or 60 points and randomly assigned to the left or right side, while the mean of the other bandit differs by 4, 8, 12, 20, or 30 points. Both reward distributions have a standard deviation of 8 points.

Each game contains either 5 or 10 trials (Fig \ref{fig:horizon}). Five-trial games correspond to the short-horizon condition (H1), whereas 10-trial games correspond to the long-horizon condition (H6).
Each game begins with four instructed trials. During these trials, participants are required to select a specified bandit and cannot choose the other option. The instructed trials control the amount of information available about each bandit before participants make their first free choice. In the unequal-information condition ([1 3]), one bandit is sampled once and the other three times. In the equal-information condition ([2 2]), both bandits are sampled twice.

Following the instructed trials, participants make either one or six free choices, depending on the horizon condition. In H1, participants make a single free choice before the game ends. In H6, participants make six free choices. Because information acquired through exploration can influence subsequent choices only in H6, exploration has greater potential value in the long-horizon condition. Our primary analyses focus on the first free-choice trial, where the information available before choice is matched across horizon conditions.

\subsection{Data set}
We aggregated data from multiple studies using the Horizon Task, with a total of 1,392 unique participants (714 females, 650 males; age range = 10–74) completing 169,376 games (84,613 H1 and 84,763 H6). Table \ref{tab:demographic} provides a detailed breakdown of demographic information, where each age group is defined as: teens ($\leq$17 years), young adults (18–34 years), midlife adults (35–59 years), and older adults ($\geq$60 years). All participants were recruited in person. No participants were excluded.

\begin{table}[h!]
\caption{Participant demographics across datasets stratified by age group}\label{tab:demographic}
\begin{tabular*}{\textwidth}{@{\extracolsep\fill}lclccccccc}
\toprule
& \multicolumn{1}{c}{Total} & & \multicolumn{3}{c}{Gender} & \multicolumn{3}{c}{Age} & \multicolumn{1}{c}{Location} \\
\cmidrule(lr){2-2}\cmidrule(lr){4-6}\cmidrule(lr){7-9}\cmidrule(lr){10-10}
Paper & $N$ & Group & F & M & O & $n$ & Mean & SD & City, State \\
\midrule
\multirow{4}{*}{\citet{Harms2024}} & \multirow{4}{*}{123} & Teens & 25 & 29 & 0 & 54 & 11.31 & 1.10 & \multirow{4}{*}{Madison, WI} \\
& & Young & 14 & 43 & 0 & 57 & 19.33 & 2.63 & \\
& & Midlife & 1 & 1 & 0 & 2 & 37.50 & 0.71 & \\
& & Unknown & 0 & 0 & 10 & 10 & -- & -- & \\
\cmidrule(lr){1-10}
\multirow{2}{*}{\begin{tabular}{@{}l@{}}\citet{Phelps2025}\\\citet{Mizell2024}\end{tabular}} & \multirow{2}{*}{349} & Young & 84 & 65 & 2 & 151 & 21.43 & 2.87 & \multirow{2}{*}{Tucson, AZ} \\
& & OA & 117 & 80 & 1 & 198 & 69.17 & 2.71 & \\
\cmidrule(lr){1-10}
\multirow{4}{*}{\citet{Sadeghiyeh2020}} & \multirow{4}{*}{369} & Teens & 3 & 3 & 0 & 6 & 17.00 & 0.00 & \multirow{4}{*}{Tucson, AZ} \\
& & Young & 105 & 234 & 1 & 340 & 19.46 & 1.57 & \\
& & Midlife & 1 & 0 & 0 & 1 & 50.00 & -- & \\
& & Unknown & 2 & 8 & 12 & 22 & -- & -- & \\
\cmidrule(lr){1-10}
\multirow{2}{*}{\citet{Smith2022}} & \multirow{2}{*}{421} & Young & 286 & 106 & 0 & 392 & 22.70 & 4.29 & \multirow{2}{*}{Tucson, AZ} \\
& & Midlife & 18 & 11 & 0 & 29 & 37.34 & 1.93 & \\
\cmidrule(lr){1-10}
\multirow{2}{*}{\begin{tabular}{@{}l@{}}\citet{Somerville2017}\\\citet{Wilson2014}\end{tabular}} & \multirow{2}{*}{92} & Teens & 47 & 44 & 0 & 91 & 14.82 & 1.76 & \multirow{2}{*}{\begin{tabular}{@{}l@{}}Cambridge, MA\\Princeton, NJ\end{tabular}} \\
& & Unknown & 0 & 1 & 0 & 1 & -- & -- & \\
\cmidrule(lr){1-10}
\multirow{3}{*}{\citet{Waltz2020}} & \multirow{3}{*}{38} & Young & 6 & 12 & 0 & 18 & 26.72 & 4.43 & \multirow{3}{*}{Baltimore, MD} \\
& & Midlife & 5 & 13 & 0 & 18 & 45.44 & 4.36 & \\
& & Unknown & 0 & 0 & 2 & 2 & -- & -- & \\
\cmidrule(lr){1-10}
\multirow{6}{*}{\textbf{Total}} & \multirow{6}{*}{\textbf{1392}} & Teens & 75 & 76 & 0 & 151 & 13.66 & 2.35 & \multirow{6}{*}{--} \\
& & Young & 495 & 460 & 3 & 958 & 21.22 & 3.64 & \\
& & Midlife & 25 & 25 & 0 & 50 & 40.52 & 5.08 & \\
& & OA & 117 & 80 & 1 & 198 & 69.17 & 2.71 & \\
& & Unknown & 2 & 9 & 24 & 35 & -- & -- & \\
& & \textbf{All} & \textbf{714} & \textbf{650} & \textbf{28} & \textbf{1392} & \textbf{--} & \textbf{--} & \\
\botrule
\end{tabular*}
\end{table}

\subsection{Baseline Model}
We used the logistic model from \citet{Wilson2014} as our baseline model. Let $Q(a)$ denote the value of action $a$, where the optimal policy selects the action with the highest value. The value of choosing option $a$ on the first free-choice trial is
\[
Q(a) = R_a + \alpha I_a + B_a,
\]
where $R_a$ is the observed mean reward of option $a$ from the instructed trials, $I_a$ encodes the uncertainty condition, and $B_a$ represents a spatial bias.

To compute the probability of choosing option $a$ over option $b$, we calculated the difference in choice values and passed it through a logistic function:

\[
p(\text{choose a}) = 
\sigma\!\left( \frac{Q(a) - Q(b)}{\eta_d} \right)=
\frac{1}{
1 + \exp\left(
-\frac{\Delta R + \alpha \Delta I + B}{\eta_d}
\right)
}
\]

where $\sigma(\cdot)$ denotes the logistic sigmoid function. $\eta_d$ is the decision noise parameter associated with random exploration. $\Delta R = R_a - R_b$ represents the difference in observed mean rewards. $\Delta I = +1$ when option $a$ is more informative, $\Delta I = -1$ when option $b$ is more informative, and $\Delta I = 0$ in the equal uncertainty condition. 

The logistic model contains three free parameters: the information bonus $\alpha$, spatial bias $B$, and decision noise $\eta_d$. We fitted the model differently depending on the level of analysis. First, following the original Horizon Task analysis \cite{Wilson2014}, the parameters were estimated separately for each participant and horizon condition. Subject-level choice parameter estimates were then averaged to obtain group-level choice curves and parameter summaries for H1 and H6. This analysis allowed us to confirm the established Horizon Task effects in the aggregated dataset. We compared parameter estimates between horizon conditions using paired t-tests. Each comparison included only participants with complete data for the relevant conditions, resulting in different sample sizes and degrees of freedom across tests. We additionally fitted one set of parameters to the pooled population and one set of parameters using only the first game played by each participant. In both cases, the model was fitted separately for the H1 and H6 conditions.

\subsection{Neural Networks}
We used neural networks to test whether they explained more behavioral variability than the baseline model. More importantly, we asked whether this additional model capacity reduced the predictability gap between H1 and H6. First, we trained models on the pooled population without subject identity. We then included subject identity to predict held-out games from participants represented during training. To test whether the gap was present before participants accumulated experience across repeated games, we repeated the analysis using only each participant's first game. Next, we incorporated cross-game history to test whether previous games provided useful behavioral context. Finally, we examined how predictability changed across the free-choice trials in H6. The data splitting procedures for these analyses are described below.

\subsubsection{Data Processing}
Model inputs were organized into static and sequential features. Static features encoded the horizon and uncertainty conditions. Sequential features consisted of choice--reward pairs $(C_t, R_t)$ from the four instructed trials, where $t \in {0,1,2,3}$ indexes the trial within a game. Sequential inputs were normalized using statistics computed from the training set. The same transformation was then applied to the validation and test sets to prevent information leakage. The architectures differed in how they combined the static and sequential features, as described in the following section.

All neural network analyses used an 80/10/10 train--validation--test split with a fixed random seed of 42. The split differed between the global and subject-level analyses. For the global models, data were split across the pooled dataset, and the training set was randomly shuffled. These models treated participants as samples from a pooled population and evaluated prediction across the general dataset. For subject-level modeling, the split was applied within each participant. Specifically, 80\% of each participant's games were used for training, 10\% for validation, and the remaining 10\% for test evaluation. We encoded subject identity as an additional static feature using one-hot encoding. For the cross-game analysis, games were ordered chronologically within each participant. 

\subsubsection{Implementation}
We trained the neural networks using the Adam optimizer \cite{Kingma2014} with a learning rate of 0.001 and a batch size of 32. Training batches were shuffled at each epoch, and the models minimized cross-entropy loss for a maximum of 100 epochs. For the cross-game models, chronological game order was preserved within each participant; only independent participant sequences were shuffled during training. At each epoch, the model updated its parameters on the training set and was then evaluated on the validation set.

We applied early stopping with a patience of five epochs. Whenever validation loss improved, we saved a copy of the model parameters from that epoch as the current best checkpoint. After early stopping, the checkpoint associated with the lowest validation loss was restored. This ensured that any decline during the patience window did not affect the final evaluation. The restored model was then evaluated once on the held-out test set. No additional regularization, such as dropout, was used because preliminary experiments showed reduced validation performance.

\subsubsection{Architectures}
We primarily evaluated recurrent neural networks (RNNs) \cite{Cho2014, Hochreiter1997} and Transformer-based architectures \cite{Vaswani2017}. We also implemented a three-layer multilayer perceptron (MLP) as a feedforward baseline to compare predictive performance across model classes. The architectures followed a hierarchy based on the behavioral information available to prediction. The MLP, base recurrent models, and attention-based models used information from the current game to capture nonlinear relationships and sequential dependencies among choices, rewards, and task conditions. We then extended the recurrent models to include history from previous games. By propagating the hidden state from one game to the next, these models could test whether influences from earlier games contributed to the unexplained variability. This addressed a limitation of the repeated-game approach, which controlled the immediate choice problem but did not account for cross-game history \cite{Wang2026}. We tested several conventional hidden-unit configurations and selected architectures that produced stable performance. Across these configurations, predictive accuracy differed by less than 1\%.

\paragraph{MLP} 

\begin{figure}[p]
\centering
  \includegraphics[width=0.7\columnwidth]{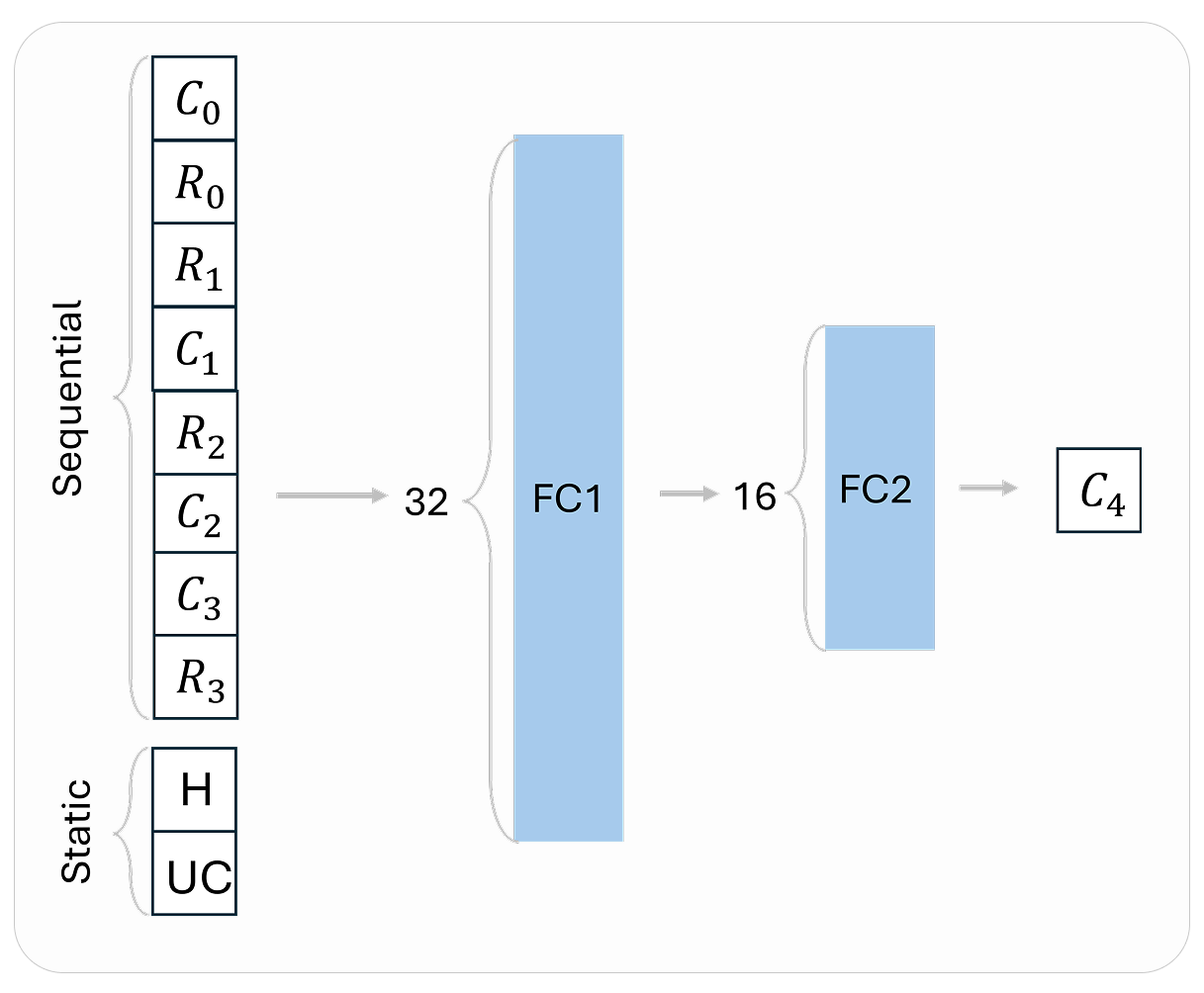}
  \caption{MLP Architecture. The model received a single concatenated vector of sequential and static inputs. Sequential inputs consisted of choice--reward pairs $(C_t, R_t)$ from the instructed trials ($t \in \{0,1,2,3\}$), while static inputs encoded horizon length (H) and uncertainty condition (UC). The architecture consisted of two FC layers (32 and 16 units, respectively) with ReLU activations. The output layer applied a softmax function to generate choice probabilities for the first free-choice trial. $C_4$ denotes the first free-choice response, as trial indexing began at 0.}~\label{fig:mlp}
\end{figure}

The MLP architecture concatenated sequential and static inputs into a single vector and predicted the binary choice on the first free-choice trial (Fig.~\ref{fig:mlp}). The network consisted of two fully connected (FC) hidden layers. A final linear layer generated logits over the two choice options, followed by a softmax function to estimate choice probabilities.

\paragraph{RNN based Models} 

\begin{figure}[p]
\centering
  \includegraphics[width=0.8\columnwidth]{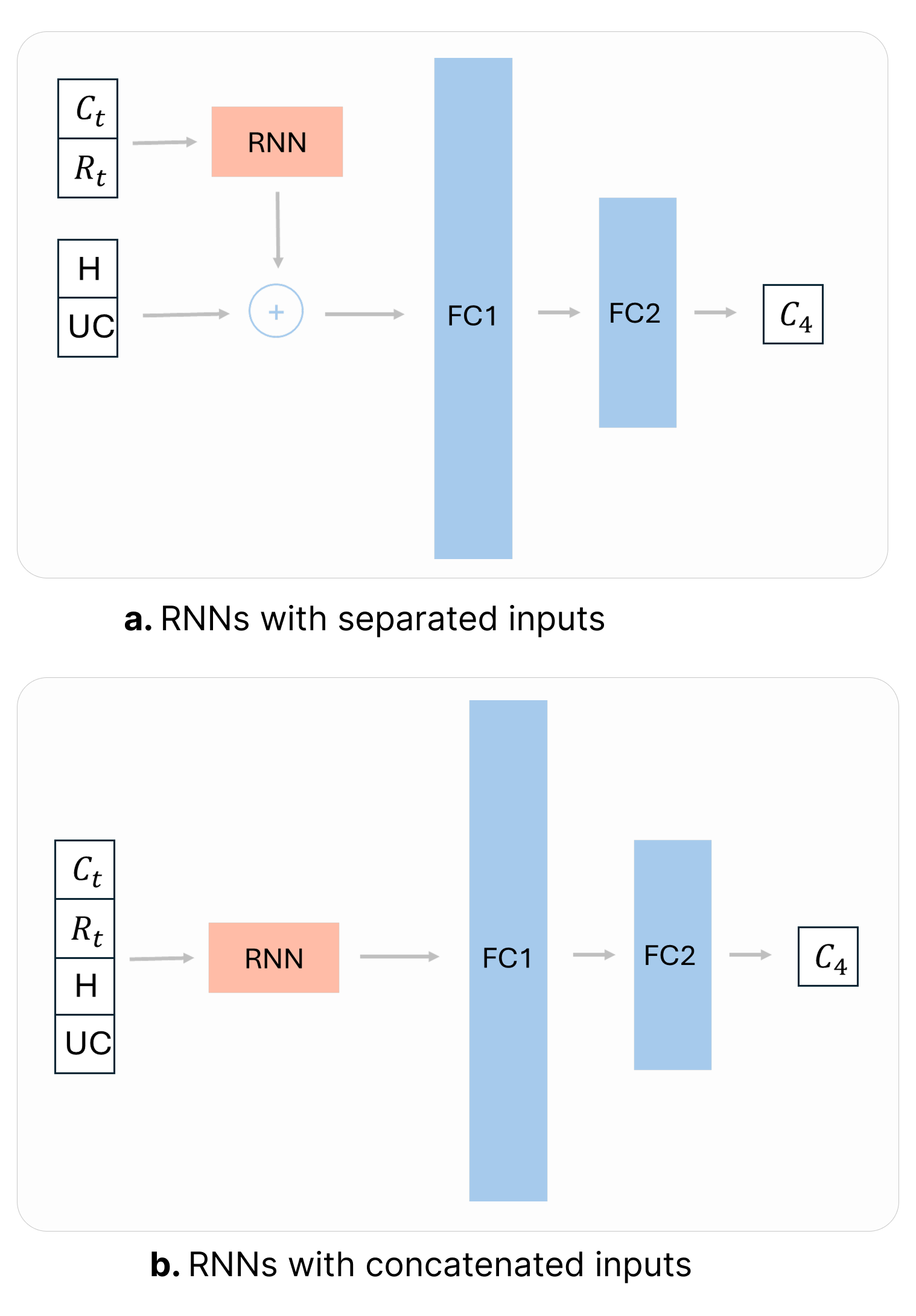}
  \caption{RNN-based Architectures. The vanilla RNN and GRU used 64 hidden units, whereas the LSTM used 128 hidden units. The TinyGRU variant had two hidden units. 
(a) The recurrent layer received only sequential inputs. The final hidden state was concatenated with static inputs. 
(b) The recurrent layer received both sequential inputs and static inputs at each time step. 
(a, b) In both architectures, the resulting representation was passed through two FC layers, identical to those used in the MLP architecture, to generate a softmax prediction of the first free-choice response.}~\label{fig:rnn}
\end{figure}

To capture the sequential structure of choices and rewards within each game, we evaluated several recurrent architectures, including vanilla RNNs, Gated Recurrent Units (GRUs) \cite{Cho2014}, and Long Short-Term Memory (LSTM) networks \cite{Hochreiter1997}. In the primary configuration (Fig.~\ref{fig:rnn}a), the recurrent layer processed only the sequential inputs. The final hidden representation was concatenated with the static inputs and passed through an FC layer before the output layer. Inspired by \citet{Ji-An2025}, we also included a TinyGRU variant with only two hidden units to establish a lower bound for representational capacity.

To test whether a participant's previous games provided useful behavioral context, we developed a cross-game variant in which the recurrent hidden state was carried over across games within each subject (Fig.~\ref{fig:rnn}b). In this variant, static inputs were integrated directly into the recurrent layer at each timestep, and the hidden state from game $g-1$ was used to initialize the state of game $g$. The importance of game history was modulated through a learnable parameter $\beta \in [0, 1]$ initialized at 0.5:
\[
h_0^{(g)} = \beta \, h_n^{(g-1)} + (1 - \beta) \, h_0^{\text{init}}.
\]
This allowed the model to capture cross-game behavioral dependencies within each subject. Each game gave the four instructed choice–reward pairs, horizon, and uncertainty condition. The state carried between games therefore summarized prior games’ instructed phases.

\paragraph{Transformer-Based Models}

\begin{figure}[p]
\centering
  \includegraphics[width=0.8\columnwidth]{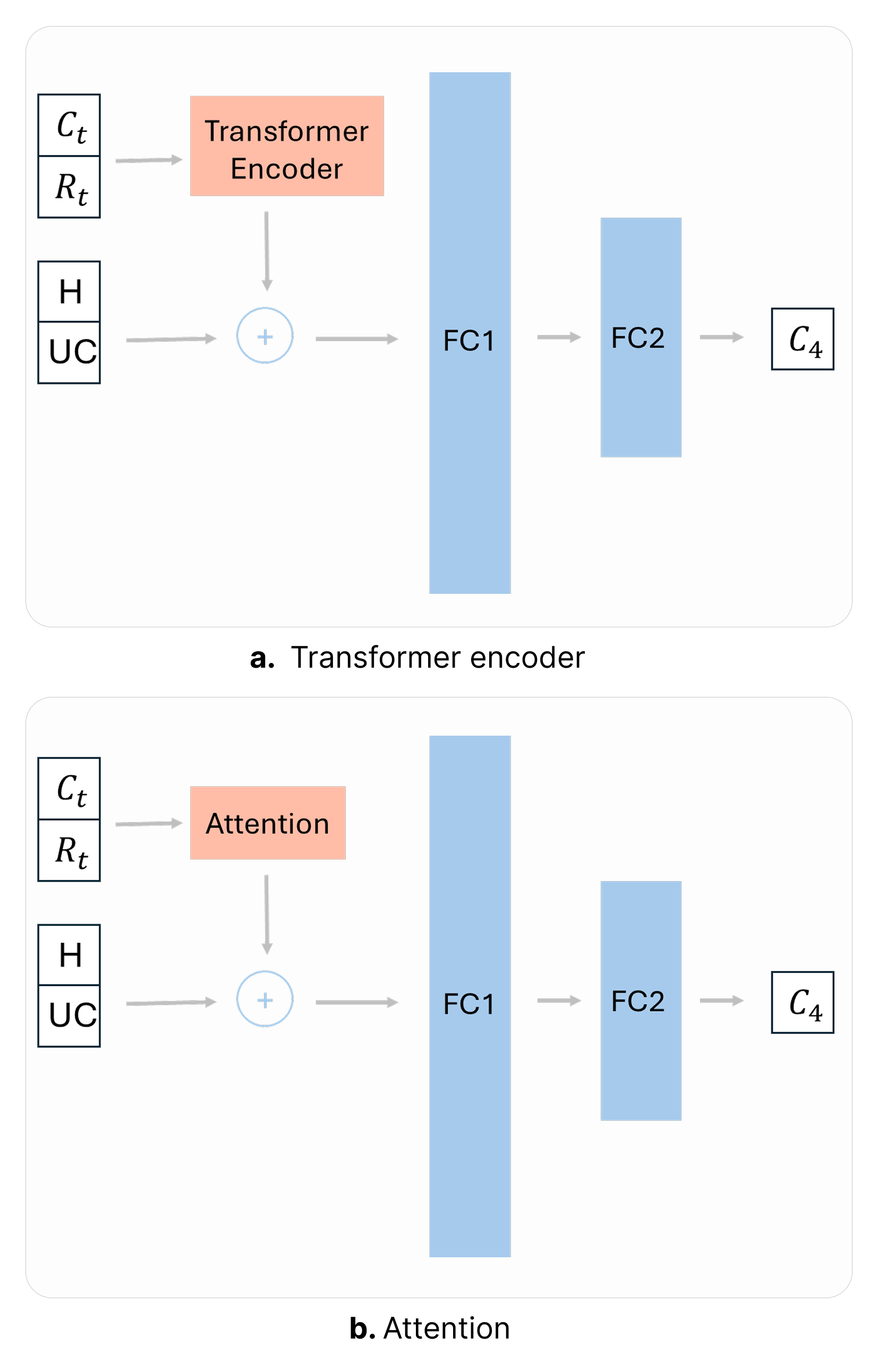}
  \caption{Transformer-based Architectures. Sequential inputs were first projected into a 64-dimensional representation with sinusoidal positional encoding.
(a) The Transformer encoder had four attention heads and a feedforward network with 128 hidden units using ReLU activation.
(b) The attention-only variant consisted of eight attention heads.
}~\label{fig:transformers}
\end{figure}

We also explored a Transformer encoder and a simplified self-attention variant (Fig.~\ref{fig:transformers}). Sequential inputs were projected into a 64-dimensional embedding space, after which sinusoidal positional encodings were added \cite{Vaswani2017}. In both architectures, the representation from the final timestep was concatenated with static inputs and passed through an FC layer before classification.

\paragraph{Seq2Seq GRU Model}

\begin{figure}[p]
\centering
  \includegraphics[width=0.8\columnwidth]{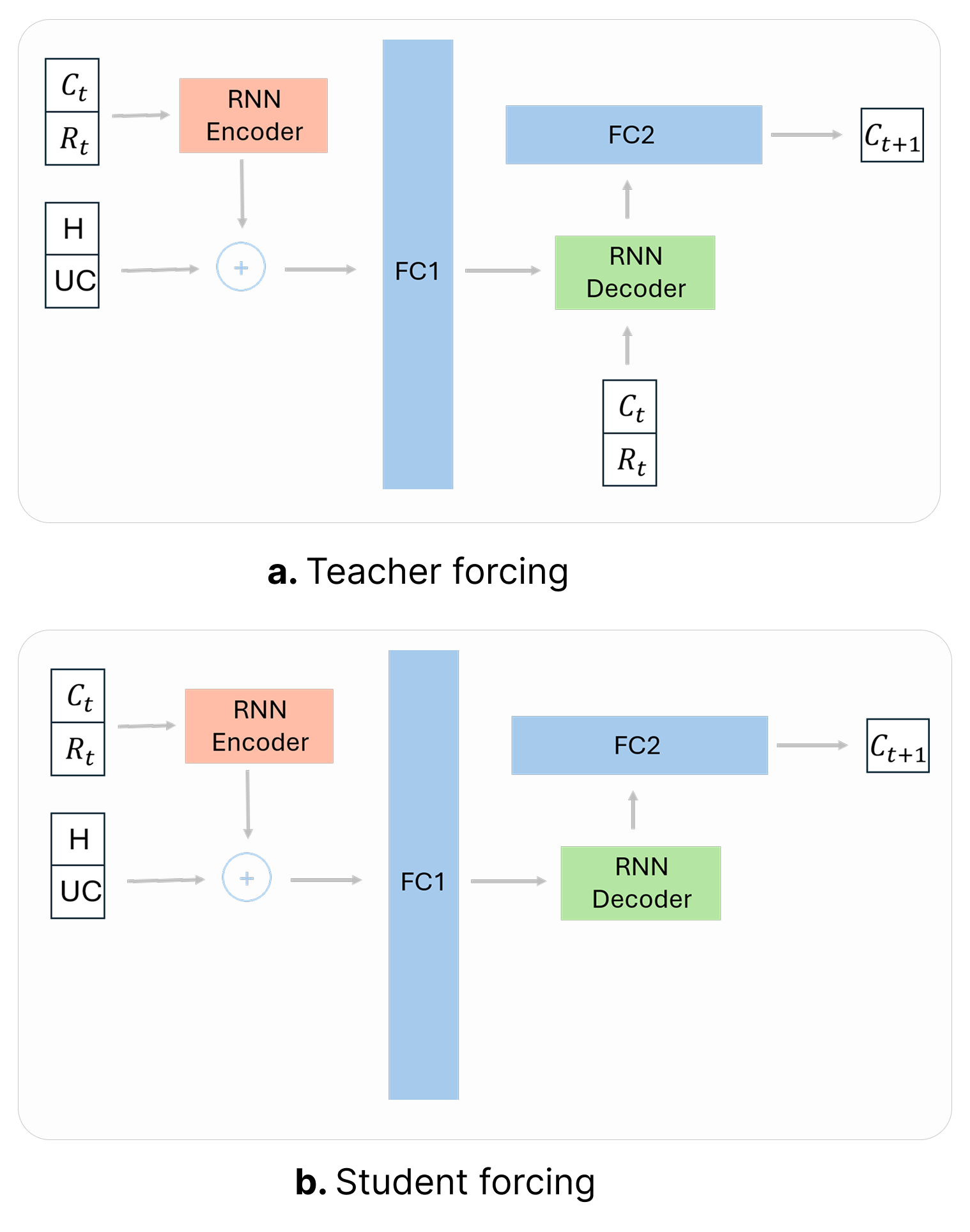}
  \caption{Seq2Seq Architecture. The model implemented an encoder--decoder recurrent framework. 
(a) In the teacher-forcing variant, the decoder received the ground-truth choice at each time step during training to predict the subsequent choice ($t \in \{4, \dots, 9\}$).
(b) In the student-forcing variant, the decoder used its own previous prediction as input to generate the next choice.}~\label{fig:seq}
\end{figure}

To examine the evolution of predictive accuracy across the six free-choice trials in the H6 condition, we implemented a sequence-to-sequence (Seq2Seq) GRU framework \cite{Sutskever2014}. The encoder processed the four instructed trials to produce a context vector. This vector was concatenated with static features and projected through a linear layer to initialize the decoder. The decoder then autoregressively generated predicted choices for the subsequent trials. We used teacher forcing during training and student forcing during evaluation to prevent ground-truth leakage (Fig.~\ref{fig:seq}). This architecture allowed us to track predictive accuracy across the H6 free-choice trials. It also allowed us to test whether performance on the final H6 trial converged toward the H1 condition.

\subsection{Evaluation}
We used multiple metrics to evaluate model performance. First, we quantified model fit using negative log-likelihood (NLL) and McFadden's pseudo-$R^2$. The null model predicted a constant probability of choosing right based on the empirical choice frequency. The logistic regression model provided the baseline NLL. For neural networks, NLL was the held-out test loss from the checkpoint selected by lowest validation loss. For each comparison, the logistic model was fitted by minimizing negative log-likelihood on the neural networks' 80\% training fold and evaluated on the identical held-out 10\% test games; the intermediate 10\% validation fold was unused by the logistic fit. To compare model fit between the neural networks and the baseline logistic model, we used NN models with encoded subject identities. Both analyses predict decisions at the subject level, rather than generalizing across the pooled population. McFadden's pseudo-$R^2$ was then computed relative to the null model. 

Second, we evaluated predictive accuracy for each model and examined model convergence across training epochs. For fair comparison with the NNs, the logistic regression baseline used out-of-sample accuracy. We then computed predictive accuracy separately for the H1 and H6 conditions.

\backmatter

% \bmhead{Supplementary information}

\bmhead{Acknowledgments}

R01AG061888 to RCW
R56 AG061888
Arizona Alzheimer’s Disease Core Center pilot grant (P30 AG019610)

\begin{appendices}

% \section{}\label{secA1}

% \begin{figure}[H]
% \centering
%   \includegraphics[width=1\columnwidth]{figures/loss.png}
% \caption{Validation loss across training epochs for neural network models in each analysis. Curves show validation cross-entropy loss during training for the global, first-game, subject-level, and cross-game comparisons. Open circles indicate the minimum validation loss for each model, corresponding to the checkpoint selected for final test-set evaluation.}~\label{fig:loss}
% \end{figure}

% \begin{figure}[H]
% \centering
%   \includegraphics[width=1\columnwidth]{figures/seq2seq_expanded.png}
%   \caption{RNN Seq2Seq Detail Architecture}~\label{fig:seq2seq_expanded}
% \end{figure}

%%=============================================%%
%% For submissions to Nature Portfolio Journals %%
%% please use the heading ``Extended Data''.   %%
%%=============================================%%

%%=============================================================%%
%% Sample for another appendix section			       %%
%%=============================================================%%

%% \section{Example of another appendix section}\label{secA2}%
%% Appendices may be used for helpful, supporting or essential material that would otherwise 
%% clutter, break up or be distracting to the text. Appendices can consist of sections, figures, 
%% tables and equations etc.

\end{appendices}

%%===========================================================================================%%
%% If you are submitting to one of the Nature Portfolio journals, using the eJP submission   %%
%% system, please include the references within the manuscript file itself. You may do this  %%
%% by copying the reference list from your .bbl file, paste it into the main manuscript .tex %%
%% file, and delete the associated \verb+\bibliography+ commands.                            %%
%%===========================================================================================%%

\bibliography{sn-bibliography}% common bib file

@article{Wilson2014,
   author = {Robert C. Wilson and Andra Geana and John M. White and Elliot A. Ludvig and Jonathan D. Cohen},
   doi = {10.1037/a0038199},
   issn = {1939-2222},
   issue = {6},
   journal = {Journal of Experimental Psychology: General},
   pages = {2074-2081},
   publisher = {American Psychological Association (APA)},
   title = {Humans use directed and random exploration to solve the explore–exploit dilemma.},
   volume = {143},
   url = {https://doi.apa.org/doi/10.1037/a0038199},
   year = {2014}
}

@article{Ji-An2025,
   author = { Ji-An Li and Marcus K. Benna and Marcelo G. Mattar},
   doi = {10.1038/s41586-025-09142-4},
   issn = {14764687},
   journal = {Nature},
   publisher = {Nature Research},
   title = {Discovering cognitive strategies with tiny recurrent neural networks},
   year = {2025}
}

@article{Wilson2021,
   author = {Robert C. Wilson and Elizabeth Bonawitz and Vincent D. Costa and R. Becket Ebitz},
   doi = {10.1016/J.COBEHA.2020.10.001},
   issn = {2352-1546},
   journal = {Current Opinion in Behavioral Sciences},
   pages = {49-56},
   publisher = {Elsevier},
   title = {Balancing exploration and exploitation with information and randomization},
   volume = {38},
   year = {2021}
}

@article{Somerville2017,
   author = {Leah H. Somerville and Stephanie F. Sasse and Megan C. Garrad and Andrew T. Drysdale and Nadine Abi Akar and Catherine Insel and Robert C. Wilson},
   doi = {10.1037/XGE0000250},
   issn = {1939-2222},
   issue = {2},
   journal = {Journal of experimental psychology. General},
   month = {2},
   pages = {155-164},
   pmid = {27977227},
   publisher = {J Exp Psychol Gen},
   title = {Charting the expansion of strategic exploratory behavior during adolescence},
   volume = {146},
   url = {https://pubmed.ncbi.nlm.nih.gov/27977227/},
   year = {2017}
}

@article{Waltz2020,
   author = {James A. Waltz and Robert C. Wilson and Matthew A. Albrecht and Michael J. Frank and James M. Gold},
   doi = {10.1162/CPSY_A_00027},
   issn = {23796227},
   journal = {Computational psychiatry (Cambridge, Mass.)},
   pages = {18},
   pmid = {33768158},
   publisher = {Web Portal Ubiquity Press},
   title = {Differential Effects of Psychotic Illness on Directed and Random Exploration},
   volume = {4},
   url = {https://pmc.ncbi.nlm.nih.gov/articles/PMC7990386/},
   year = {2020}
}

@article{Smith2022,
   author = {Ryan Smith and Samuel Taylor and Robert C. Wilson and Anne E. Chuning and Michelle R. Persich and Siyu Wang and William D.S. Killgore},
   doi = {10.3389/fpsyt.2021.782136},
   issn = {16640640},
   journal = {Frontiers in Psychiatry},
   month = {1},
   publisher = {Frontiers Media S.A.},
   title = {Lower Levels of Directed Exploration and Reflective Thinking Are Associated With Greater Anxiety and Depression},
   volume = {12},
   year = {2022}
}

@article{Sadeghiyeh2020,
   author = {Hashem Sadeghiyeh and Siyu Wang and Maxwell R. Alberhasky and Hannah M. Kyllo and Amitai Shenhav and Robert C. Wilson},
   doi = {10.1038/s41598-020-60576-4},
   issn = {20452322},
   issue = {1},
   journal = {Scientific Reports},
   month = {12},
   pmid = {32132573},
   publisher = {Nature Research},
   title = {Temporal discounting correlates with directed exploration but not with random exploration},
   volume = {10},
   year = {2020}
}

@article{Feng2021,
   author = {Samuel F. Feng and Siyu Wang and Sylvia Zarnescu and Robert C. Wilson},
   doi = {10.1038/s41598-021-82530-8},
   issn = {20452322},
   issue = {1},
   journal = {Scientific Reports},
   pmid = {33542333},
   publisher = {Nature Research},
   title = {The dynamics of explore–exploit decisions reveal a signal-to-noise mechanism for random exploration},
   volume = {11},
   year = {2021}
}

@article{Harms2024,
   author = {Madeline B. Harms and Yuyan Xu and C. Shawn Green and Kristina Woodard and Robert Wilson and Seth D. Pollak},
   doi = {10.1016/j.cogpsych.2024.101650},
   issn = {00100285},
   journal = {Cognitive Psychology},
   month = {5},
   pmid = {38461609},
   publisher = {Academic Press Inc.},
   title = {The structure and development of explore-exploit decision making},
   volume = {150},
   year = {2024}
}

@article{Phelps2025,

   author = {Caroline E. Phelps and Alec E. Frisvold and Meghna Sreeram and Victoria Antoniou and Larissa E. Oliveira and Sierra R. Tooke and Yan Z. Lu and Anna C. Lemon and Joseph U. Fraire and Alexe G. Delval and Betsy K. Smith and Devynn H. Spangenberg and Madelien N. Mithelavage and Michelle N. Ngo and Elizabeth MO. Keller and Liliana J. Issac and Sara A. Harader and Jack-Morgan Mizell and Siyu Wang and Waitsang Keung and Mark H. Sundman and Mary-Kathryn Franchetti and Ying-hui Chou and Gene E. Alexander and Robert C. Wilson},
   doi = {10.1101/2025.02.13.638023},
   institution = {bioRxiv},
   journal = {bioRxiv},
   month = {2},
   pages = {2025.02.13.638023},
   publisher = {Cold Spring Harbor Laboratory},
   title = {The dynamics of explore-exploit decisions suggest a threshold mechanism for reduced random exploration in older adults},
   url = {https://www.biorxiv.org/content/10.1101/2025.02.13.638023v1 https://www.biorxiv.org/content/10.1101/2025.02.13.638023v1.abstract},
   year = {2025}
}

@article{Mizell2024,

   author = {Jack Morgan Mizell and Siyu Wang and Alec Frisvold and Lily Alvarado and Alex Farrell-Skupny and Waitsang Keung and Caroline E. Phelps and Mark H. Sundman and Mary Kathryn Franchetti and Ying hui Chou and Gene E. Alexander and Robert C. Wilson},
   doi = {10.1037/PAG0000791},
   issn = {1939-1498},
   issue = {1},
   journal = {Psychology and aging},
   pages = {88-101},
   pmid = {38358695},
   publisher = {Psychol Aging},
   title = {Differential impacts of healthy cognitive aging on directed and random exploration},
   volume = {39},
   url = {https://pubmed.ncbi.nlm.nih.gov/38358695/},
   year = {2024}
}

@article{Kingma2014,
   author = {Diederik P. Kingma and Jimmy Lei Ba},
   journal = {3rd International Conference on Learning Representations, ICLR 2015 - Conference Track Proceedings},
   month = {12},
   publisher = {International Conference on Learning Representations, ICLR},
   title = {Adam: A Method for Stochastic Optimization},
   url = {https://arxiv.org/pdf/1412.6980},
   year = {2014}
}

@article{Vaswani2017,
  title={Attention is all you need},
  author={Vaswani, Ashish and Shazeer, Noam and Parmar, Niki and Uszkoreit, Jakob and Jones, Llion and Gomez, Aidan N and Kaiser, {\L}ukasz and Polosukhin, Illia},
  journal={Advances in neural information processing systems},
  volume={30},
  year={2017}
}

@article{Cho2014,
   author = {Kyunghyun Cho and Bart Van Merriënboer and Caglar Gulcehre and Dzmitry Bahdanau and Fethi Bougares and Holger Schwenk and Yoshua Bengio},
   doi = {10.3115/v1/d14-1179},
   isbn = {9781937284961},
   journal = {EMNLP 2014 - 2014 Conference on Empirical Methods in Natural Language Processing, Proceedings of the Conference},
   month = {6},
   pages = {1724-1734},
   publisher = {Association for Computational Linguistics (ACL)},
   title = {Learning Phrase Representations using RNN Encoder-Decoder for Statistical Machine Translation},
   url = {https://arxiv.org/pdf/1406.1078},
   year = {2014}
}

@article{Hochreiter1997,
  title={Long short-term memory},
  author={Hochreiter, Sepp and Schmidhuber, Jürgen},
  journal={Neural computation},
  volume={9},
  number={8},
  pages={1735--1780},
  year={1997},
  publisher={MIT Press}
}

@article{Sutskever2014,
   author = {Ilya Sutskever and Oriol Vinyals and Quoc V. Le},
   issn = {10495258},
   issue = {January},
   journal = {Advances in Neural Information Processing Systems},
   month = {9},
   pages = {3104-3112},
   publisher = {Neural information processing systems foundation},
   title = {Sequence to Sequence Learning with Neural Networks},
   volume = {4},
   url = {https://arxiv.org/pdf/1409.3215},
   year = {2014}
}

@article{Sutton1998,
  author={Sutton, R.S. and Barto, A.G.},
  journal={IEEE Transactions on Neural Networks}, 
  title={Reinforcement Learning: An Introduction}, 
  year={1998},
  volume={9},
  number={5},
  pages={1054-1054},
  doi={10.1109/TNN.1998.712192}}

@article{kaelbling1996reinforcementlearningsurvey,
      title={Reinforcement Learning: A Survey}, 
      author={L. P. Kaelbling and M. L. Littman and A. W. Moore},
      year={1996},
      eprint={cs/9605103},
      archivePrefix={arXiv},
      primaryClass={cs.AI},
      url={https://arxiv.org/abs/cs/9605103}, 
}

@article{bellman1956problem,
  title={A problem in the sequential design of experiments},
  author={Bellman, Richard},
  journal={Sankhy{\=a}: The Indian Journal of Statistics (1933-1960)},
  volume={16},
  number={3/4},
  pages={221--229},
  year={1956},
  publisher={JSTOR}
}

@article{gittins1979bandit,
  title={Bandit processes and dynamic allocation indices},
  author={Gittins, John C},
  journal={Journal of the Royal Statistical Society Series B: Statistical Methodology},
  volume={41},
  number={2},
  pages={148--164},
  year={1979},
  publisher={Oxford University Press}
}

@article{zhang2013forgetful,
  title={Forgetful Bayes and myopic planning: Human learning and decision-making in a bandit setting},
  author={Zhang, Shunan and Yu, Angela J},
  journal={Advances in neural information processing systems},
  volume={26},
  year={2013}
}

@article{averbeck2015theory,
  title={Theory of choice in bandit, information sampling and foraging tasks},
  author={Averbeck, Bruno B},
  journal={PLoS computational biology},
  volume={11},
  number={3},
  pages={e1004164},
  year={2015},
  publisher={Public Library of Science San Francisco, CA USA}
}

@article{Beck2012,
   author = {Jeffrey M. Beck and Wei Ji Ma and Xaq Pitkow and Peter E. Latham and Alexandre Pouget},
   doi = {10.1016/j.neuron.2012.03.016},
   issn = {08966273},
   issue = {1},
   journal = {Neuron},
   month = {4},
   pages = {30-39},
   pmid = {22500627},
   publisher = {Elsevier},
   title = {Not Noisy, Just Wrong: The Role of Suboptimal Inference in Behavioral Variability},
   volume = {74},
   url = {https://www.cell.com/action/showFullText?pii=S0896627312002802 https://www.cell.com/action/showAbstract?pii=S0896627312002802 https://www.cell.com/neuron/abstract/S0896-6273(12)00280-2},
   year = {2012}
}

@article{osborne2005sensory,
  title={A sensory source for motor variation},
  author={Osborne, Leslie C and Lisberger, Stephen G and Bialek, William},
  journal={Nature},
  volume={437},
  number={7057},
  pages={412--416},
  year={2005},
  publisher={Nature Publishing Group UK London}
}

@article{faisal2008noise,
  title={Noise in the nervous system},
  author={Faisal, A Aldo and Selen, Luc PJ and Wolpert, Daniel M},
  journal={Nature reviews neuroscience},
  volume={9},
  number={4},
  pages={292--303},
  year={2008},
  publisher={Nature Publishing Group UK London}
}

@article{Tervo2014,
   author = {Dougal G.R. Tervo and Mikhail Proskurin and Maxim Manakov and Mayank Kabra and Alison Vollmer and Kristin Branson and Alla Y. Karpova},
   doi = {10.1016/j.cell.2014.08.037},
   issn = {10974172},
   issue = {1},
   journal = {Cell},
   month = {9},
   pages = {21-32},
   pmid = {25259917},
   publisher = {Cell Press},
   title = {Behavioral variability through stochastic choice and its gating by anterior cingulate cortex},
   volume = {159},
   url = {https://www.cell.com/action/showFullText?pii=S0092867414011076 https://www.cell.com/action/showAbstract?pii=S0092867414011076 https://www.cell.com/cell/abstract/S0092-8674(14)01107-6},
   year = {2014}
}

@article{Kirkpatrick1983,
  author  = {Kirkpatrick, Scott and Gelatt, C. Daniel and Vecchi, Mario P.},
  title   = {Optimization by Simulated Annealing},
  journal = {Science},
  year    = {1983},
  volume  = {220},
  number  = {4598},
  pages   = {671--680},
  doi     = {10.1126/science.220.4598.671}
}

@article{Cerny1985,
  author  = {{\v{C}}ern{\'y}, Vladim{\'i}r},
  title   = {Thermodynamical Approach to the Traveling Salesman Problem: An Efficient Simulation Algorithm},
  journal = {Journal of Optimization Theory and Applications},
  year    = {1985},
  volume  = {45},
  number  = {1},
  pages   = {41--51},
  doi     = {10.1007/BF00940812}
}

@article{Geman1984,
  author  = {Geman, Stuart and Geman, Donald},
  title   = {Stochastic Relaxation, {Gibbs} Distributions, and the {Bayesian} Restoration of Images},
  journal = {IEEE Transactions on Pattern Analysis and Machine Intelligence},
  year    = {1984},
  volume  = {PAMI-6},
  number  = {6},
  pages   = {721--741},
  doi     = {10.1109/TPAMI.1984.4767596}
}

@article{Gelfand1990,
  author  = {Gelfand, Alan E. and Smith, Adrian F. M.},
  title   = {Sampling-Based Approaches to Calculating Marginal Densities},
  journal = {Journal of the American Statistical Association},
  year    = {1990},
  volume  = {85},
  number  = {410},
  pages   = {398--409},
  doi     = {10.1080/01621459.1990.10476213}
}

@article{Metropolis1953,
  author  = {Metropolis, Nicholas and Rosenbluth, Arianna W. and Rosenbluth, Marshall N. and Teller, Augusta H. and Teller, Edward},
  title   = {Equation of State Calculations by Fast Computing Machines},
  journal = {The Journal of Chemical Physics},
  year    = {1953},
  volume  = {21},
  number  = {6},
  pages   = {1087--1092},
  doi     = {10.1063/1.1699114}
}

@article{Hastings1970,
  author  = {Hastings, W. K.},
  title   = {Monte Carlo Sampling Methods Using Markov Chains and Their Applications},
  journal = {Biometrika},
  year    = {1970},
  volume  = {57},
  number  = {1},
  pages   = {97--109},
  doi     = {10.1093/biomet/57.1.97}
}

@article{Wang2026,
   author = {Siyu Wang and Robert C. Wilson},
   doi = {10.1371/JOURNAL.PCBI.1014026},
   issn = {15537358},
   issue = {3},
   journal = {PLOS Computational Biology},
   month = {3},
   pages = {e1014026},
   pmid = {41843633},
   title = {Separating random and deterministic sources of computational noise in explore-exploit decisions},
   volume = {22},
   url = {https://pmc.ncbi.nlm.nih.gov/articles/PMC13020999/},
   year = {2026}
}

@article{krebs1978test,
  title={Test of optimal sampling by foraging great tits},
  author={Krebs, John R and Kacelnik, Alejandro and Taylor, Peter},
  journal={Nature},
  volume={275},
  number={5675},
  pages={27--31},
  year={1978},
  publisher={Nature Publishing Group UK London}
}

@misc{watkins1989learning,
  title={Learning from delayed rewards.},
  author={Watkins, CJCH and others},
  year={1989}
}

@article{gershman2018dopaminergic,
  title={Dopaminergic genes are associated with both directed and random exploration},
  author={Gershman, Samuel J and Tzovaras, Bastian Greshake},
  journal={Neuropsychologia},
  volume={120},
  pages={97--104},
  year={2018},
  publisher={Elsevier}
}

@article{schulz2019algorithmic,
  title={The algorithmic architecture of exploration in the human brain},
  author={Schulz, Eric and Gershman, Samuel J},
  journal={Current opinion in neurobiology},
  volume={55},
  pages={7--14},
  year={2019},
  publisher={Elsevier}
}

@article{gershman2018deconstructing,
  title={Deconstructing the human algorithms for exploration},
  author={Gershman, Samuel J},
  journal={Cognition},
  volume={173},
  pages={34--42},
  year={2018},
  publisher={Elsevier}
}

@article{frank2009prefrontal,
  title={Prefrontal and striatal dopaminergic genes predict individual differences in exploration and exploitation},
  author={Frank, Michael J and Doll, Bradley B and Oas-Terpstra, Jen and Moreno, Francisco},
  journal={Nature neuroscience},
  volume={12},
  number={8},
  pages={1062--1068},
  year={2009},
  publisher={Nature Publishing Group US New York}
}

@article{badre2012rostrolateral,
  title={Rostrolateral prefrontal cortex and individual differences in uncertainty-driven exploration},
  author={Badre, David and Doll, Bradley B and Long, Nicole M and Frank, Michael J},
  journal={Neuron},
  volume={73},
  number={3},
  pages={595--607},
  year={2012},
  publisher={Elsevier}
}

@article{cavanagh2012frontal,
  title={Frontal theta reflects uncertainty and unexpectedness during exploration and exploitation},
  author={Cavanagh, James F and Figueroa, Christina M and Cohen, Michael X and Frank, Michael J},
  journal={Cerebral cortex},
  volume={22},
  number={11},
  pages={2575--2586},
  year={2012},
  publisher={Oxford University Press}
}

@article{speekenbrink2015uncertainty,
  title={Uncertainty and exploration in a restless bandit problem},
  author={Speekenbrink, Maarten and Konstantinidis, Emmanouil},
  journal={Topics in cognitive science},
  volume={7},
  number={2},
  pages={351--367},
  year={2015},
  publisher={Wiley Online Library}
}

@article{daw2006cortical,
  title={Cortical substrates for exploratory decisions in humans},
  author={Daw, Nathaniel D and O'doherty, John P and Dayan, Peter and Seymour, Ben and Dolan, Raymond J},
  journal={Nature},
  volume={441},
  number={7095},
  pages={876--879},
  year={2006},
  publisher={Nature Publishing Group UK London}
}

@article{gershman2019uncertainty,
  title={Uncertainty and exploration},
  author={Gershman, Samuel J.},
  journal={Decision},
  volume={6},
  number={3},
  pages={277--286},
  year={2019},
  publisher={American Psychological Association},
  doi={10.1037/dec0000101}
}

@article{xiong2025strategy,
  title={Human Strategy Adaptation in Reinforcement Learning Resembles Policy Gradient Ascent},
  author={Xiong, Hua-Dong and Ji-An, Li and Wilson, Robert C. and Mattar, Marcelo G.},
  journal={bioRxiv},
  year={2025},
  doi={10.1101/2025.07.28.667308}
}

@article{Budzillo2017,
  title={Dopaminergic Modulation of Basal Ganglia Output through Coupled Excitation--Inhibition},
  author={Budzillo, Agata and Duffy, Alison and Miller, Kimberly E. and Fairhall, Adrienne L. and Perkel, David J.},
  journal={Proceedings of the National Academy of Sciences},
  volume={114},
  number={22},
  pages={5713--5718},
  year={2017},
  doi={10.1073/pnas.1611146114}
}

@article{Ebitz2018,
  title={Exploration Disrupts Choice-Predictive Signals and Alters Dynamics in Prefrontal Cortex},
  author={Ebitz, R. Becket and Albarran, Eddy and Moore, Tirin},
  journal={Neuron},
  volume={97},
  number={2},
  pages={450--461.e9},
  year={2018},
  doi={10.1016/j.neuron.2017.12.007}
}

@article{Findling2019,
  title={Computational Noise in Reward-Guided Learning Drives Behavioral Variability in Volatile Environments},
  author={Findling, Charles and Skvortsova, Vasilisa and Dromnelle, R{\'e}mi and Palminteri, Stefano and Wyart, Valentin},
  journal={Nature Neuroscience},
  volume={22},
  number={12},
  pages={2066--2077},
  year={2019},
  doi={10.1038/s41593-019-0518-9}
}

@article{Peterson2021,
  title={Using Large-Scale Experiments and Machine Learning to Discover Theories of Human Decision-Making},
  author={Peterson, Joshua C. and Bourgin, David D. and Agrawal, Mayank and Reichman, Daniel and Griffiths, Thomas L.},
  journal={Science},
  volume={372},
  number={6547},
  pages={1209--1214},
  year={2021},
  doi={10.1126/science.abe2629}
}

@article{Wang2025FLAG,
  title={Modeling Value Integration During Decision Making Under Uncertainty with the Florida and Georgia Gambling Task},
  author={Wang, Shensheng and Wilson, Robert C. and Ebner, Natalie C. and Lighthall, Nichole R.},
  journal={Scientific Reports},
  volume={15},
  pages={36826},
  year={2025},
  doi={10.1038/s41598-025-08333-3}
}

@article{Musall2019,
  title={Single-trial neural dynamics are dominated by richly varied movements},
  author={Musall, Simon and Kaufman, Matthew T. and Juavinett, Ashley L. and Gluf, Steven and Churchland, Anne K.},
  journal={Nature Neuroscience},
  volume={22},
  number={10},
  pages={1677--1686},
  year={2019},
  doi={10.1038/s41593-019-0502-4}
}

@article{lecun2015deep,
  title={Deep learning},
  author={LeCun, Yann and Bengio, Yoshua and Hinton, Geoffrey},
  journal={nature},
  volume={521},
  number={7553},
  pages={436--444},
  year={2015},
  publisher={Nature Publishing Group UK London}
}

@article{hornik1991approximation,
  title={Approximation capabilities of multilayer feedforward networks},
  author={Hornik, Kurt},
  journal={Neural networks},
  volume={4},
  number={2},
  pages={251--257},
  year={1991},
  publisher={Elsevier}
}

@article{Cybenko1989,
  title={Approximation by superpositions of a sigmoidal function},
  author={Cybenko, George},
  journal={Mathematics of Control, Signals and Systems},
  volume={2},
  pages={303--314},
  year={1989},
  doi={10.1007/BF02551274}
}
%% if required, the content of .bbl file can be included here once bbl is generated
%%\input sn-article.bbl

\end{document}